\documentclass[10pt,twocolumn,english,aps,pra,nofootinbib,superscriptaddress,showpacs,showkeys,longbibliography]{revtex4-1}
\usepackage[T1]{fontenc}
\usepackage{color}
\usepackage{babel}
\usepackage{multirow}
\usepackage{bbold}
\usepackage{url}
\usepackage{textcomp}
\usepackage{amssymb}
\usepackage{amsmath}
\usepackage{mathtools}
\usepackage{siunitx}
\usepackage{verbatim}
\usepackage{graphicx}
\usepackage{subfigure}
\graphicspath{ {images/} }
\usepackage{bbm, dsfont}
\usepackage{amsfonts}
\usepackage{babel}
\usepackage{mathptmx}

\definecolor{myurlcolor}{rgb}{0,0,0.7}
\definecolor{myrefcolor}{rgb}{0.8,0,0}
\usepackage[unicode=true,pdfusetitle, bookmarks=false,bookmarksnumbered=false,
bookmarksopen=false, breaklinks=false,pdfborder={0 0 0},backref=false,
colorlinks=true, linkcolor=myrefcolor,citecolor=myurlcolor,urlcolor=myurlcolor]
{hyperref}

\usepackage{hyperref}

\begin{document}

\title{Development of swarm behavior in artificial learning agents that adapt \\ to different foraging environments}
\author{Andrea L{\'o}pez-Incera}
\email{andrea.lopez-incera@uibk.ac.at}
\affiliation{Institute for Theoretical Physics, University of Innsbruck, A-6020 Innsbruck, Austria}
\author{Katja Ried}
\email{katja.ried@uibk.ac.at}
\affiliation{Institute for Theoretical Physics, University of Innsbruck, A-6020 Innsbruck, Austria}
\author{Thomas M\"uller}
\email{thomas.mueller@uni-konstanz.de}
\affiliation{Fachbereich Philosophie, Universität Konstanz, Fach 17, 78457 Konstanz, Germany}
\author{Hans J. Briegel}
\email{hans.briegel@uibk.ac.at}
\affiliation{Institute for Theoretical Physics, University of Innsbruck, A-6020 Innsbruck, Austria}
\affiliation{Fachbereich Philosophie, Universität Konstanz, Fach 17, 78457 Konstanz, Germany}
\begin{abstract}
Collective behavior, and swarm formation in particular, has been studied from several perspectives within a large variety of fields, ranging from biology to physics. In this work, we apply Projective Simulation to model each individual as an artificial learning agent that interacts with its neighbors and surroundings in order to make decisions and learn from them. Within a reinforcement learning framework, we discuss one-dimensional learning scenarios where agents need to get to food resources to be rewarded. We observe how different types of collective motion emerge depending on the distance the agents need to travel to reach the resources. For instance, strongly aligned swarms emerge when the food source is placed far away from the region where agents are situated initially. In addition, we study the properties of the individual trajectories that occur within the different types of emergent collective dynamics. Agents trained to find distant resources exhibit individual trajectories with L\'evy-like characteristics as a consequence of the collective motion, whereas agents trained to reach nearby resources present Brownian-like trajectories.
\end{abstract}
\pacs{}
\maketitle

\section{Introduction}
Collective behavior is a common but intriguing phenomenon in nature. Species as diverse as locusts, and some families of fish or birds exhibit different types of collective motion in very different environments and situations. Although the general properties of swarms, schools and flocks have been widely studied (see e.g. \citep{Vicsek12} for a review), the emergence of global, coordinated motion from the individual actions is still a subject of study. Different approaches, ranging from statistical physics to agent-based models, have led to new insights and descriptions of the phenomenon. Statistical physics models are very successful at describing macroscopic properties such as phase transitions and metastable states \citep{Yates09,Kolpas07,Bode10}, but in order to apply the powerful tools of statistical mechanics, these models normally simplify the individuals to particles that interact according to certain rules dictated by the physical model adopted, as for instance the Ising-type interaction of the spins in a lattice. A different type of models are the so-called self-propelled particle (SPP) models \citep{Vicsek95,Czirok99,Czirok3D99,Oloan99}, which enable higher complexity in descriptions at the individual level but still allow one to employ the tools of statistical physics. They describe individuals as particles that move with a constant velocity and interact with other individuals via fixed sets of rules that are externally imposed. In SPP models, the description of the interactions is not restricted to physically accepted first principles, but can include ad hoc rules based on specific experimental observations.

In this work, we follow a different approach and model the individuals as artificial learning agents. In particular, we apply Projective Simulation (PS) \citep{BriegelCuevas12}, which is a model of agency that can incorporate learning processes via a reinforcement learning mechanism. The individuals are thus described as PS agents that interact with their surroundings, make decisions accordingly and learn from them based on rewards provided by the environment. This framework allows for a more detailed, realistic description in terms of the perceptual apparatus of the agent. One of the main differences with respect to previous models is that the interaction rules between agents are not imposed or fixed in advance, but they emerge as the result of learning in a given task environment. This type of agent-based models that employ artificial intelligence to model behavior are gaining popularity in the last few years. Artificial neural networks (ANN) have been used, for instance, in the context of navigation behaviors \citep{Morales05,Mueller11} and reinforcement learning (RL) algorithms have been applied to model collective behavior in different scenarios, such as pedestrian movement \citep{Martinez17} or flocking \citep{Shimada18,Durve19}.

In contrast to other learning models such as neural networks, PS provides a transparent, explicit structure that can be analyzed and interpreted. This feature is particularly useful in modeling collective behavior, since we can study the individual decision making processes, what the agents learn and why they learn it. This way, we can directly address the questions of how and why particular individual interactions arise that in turn lead to collective behaviors. Initial work by Ried et al. \citep{Ried19}, where the authors use PS to model the density-dependent swarm behavior of locusts, laid the foundations of the present work.

Since the interaction rules are developed by the agents themselves, the challenge is to design the environment and learning task that will give rise to the individual and, consequently, collective behavior. In previous works, the agents are directly rewarded for aligning themselves with the surrounding agents \citep{Ried19} or for not losing neighbours \citep{Durve19}. Instead of rewarding a specific behavior, in this work we set a survival task that the agents need to fulfill in order to get the reward, and then analyze the emergent behavioral dynamics.

As a starting hypothesis, we consider the need to forage as an evolutionary pressure and design a learning task that consists in finding a remote food source. Due to this particular survival task, our work relates to the investigation of foraging theories and optimal searching behavior.

There is a vast number of studies devoted to the analysis of foraging strategies in different types of environments e.g., \citep{Sinervo97,Stephens07,Pyke84,vis11}. In the particular case of environments with sparsely distributed resources (e.g. patchy landscapes), there are two main candidates for the optimal search model: L\'evy walks \citep{Levy37,Shlesinger86,Viswanathan99} and composite correlated random walks (CCRW) \citep{Benhamou92,Benhamou07}. Although the mathematical models behind them are fundamentally different, they have some common features that make the movement patterns hard to distinguish \citep{Viswanathan96,Sims08,Benhamou07,Edwards11,Edwards12}. In broad terms, both models can produce trajectories that are a combination of short steps (with large turning angles in 2D), which are useful for exploring the patch area, and long, straight steps, which are efficient to travel the inter-patch distances. Even though both models have theoretical \citep{Viswanathan99,Benhamou92} and experimental (e.g. \citep{Humphries12,Dragon12}) support, it is not yet clear if animal foraging patterns can be described and explained by such models or if they are too complex to admit such simplifications. 

Due to the fact that our learning task is directly related to foraging strategies, we link the present work to the aforementioned studies by analyzing the individual trajectories the agents produce as a consequence of the behavior developed in the different learning contexts.

The paper is organized as follows: an introduction to Projective Simulation and a detailed description of the model and the learning setup are given in Sec.~\ref{SEC The model and the learning setup}. In Sec.~\ref{learning diff scen}, we present different learning tasks and analyze the resulting learned behaviors. In Sec.~\ref{SEC Analysis of the learned dynamics}, we study the emergent group dynamics and individual trajectories within the framework of search models to determine if they can be described as L\'evy walks or composite correlated random walks. Finally, we summarize the results and conclude in Sec.~\ref{SEC summary}.

\section{The model and the learning setup}\label{SEC The model and the learning setup}
A wide range of models and techniques have been applied to the study of collective behavior. In this work, we apply Projective Simulation, a model for artificial agency \citep{BriegelCuevas12,Mautner15,Makmal16,Melnikov17,Melnikov18,RiedEva19}. Each individual is an artificial agent that can perceive its surroundings, make decisions and perform actions. Within the PS model, the agent's decision making is integrated into a framework for reinforcement learning (RL) that allows one to design concrete scenarios and tasks that the individuals should solve and then study the resulting strategies\footnote{We remark that the notion of \textit{strategy} employed throughout this work does not imply that the agents are able to \textit{plan}. We use the word "strategy" to refer to the behavior the agents develop given a certain learning task.} developed by the agents. In addition, each agent's motor and sensory abilities can be modeled in a detailed, realistic way.

In our model of collective behavior, the interaction rules with other individuals are not fixed in advance; instead the agents develop them based on their previous experience and learning. The most natural interpretation of this approach is that it describes how a group of given individuals change their behavior over the course of their interactions, for example human children at play. However, our artificial learning agents can also be used to model simpler entities that do not exhibit learning in the sense of noticeable modifications of their responses over the course of a single individual's lifetime, but only change their behavior over the course of several generations. In this case, a single simulated agent does not correspond to one particular individual, in one particular generation, but rather stands as an avatar for a generic individual throughout the entire evolution of the species. The evolutionary pressures driving behavioural changes over this time-scale can be easily encoded in a RL scenario, since the reward scheme can be designed in such a way that only the behaviors that happen to be beneficial under these pressures are rewarded. This allows us to directly test whether the evolutionary pressures are a possible \textit{causal} explanation for the observed behavior or not. 

Although other reinforcement learning algorithms may be used to model a learning agent, Projective Simulation is particularly suitable for the purpose of modeling collective behavior, since it provides a clear and transparent structure that gives direct access to the internal state of the agent, so that the deliberation process can be analyzed in an explicit way and can be related to the agent's behavior. This analysis can help us gain new insight into how and why the individual interactions that lead to collective behaviors emerge.

\subsection{Projective Simulation}\label{PS}
Projective Simulation (PS) is a model for artificial agency that is based on the notion of episodic memory \citep{BriegelCuevas12}. The agent interacts with its surroundings and receives some inputs called percepts, which trigger a deliberation process that leads to the agent performing an action on the environment. 

\begin{figure}[ht!]
\includegraphics[width=3.4in]{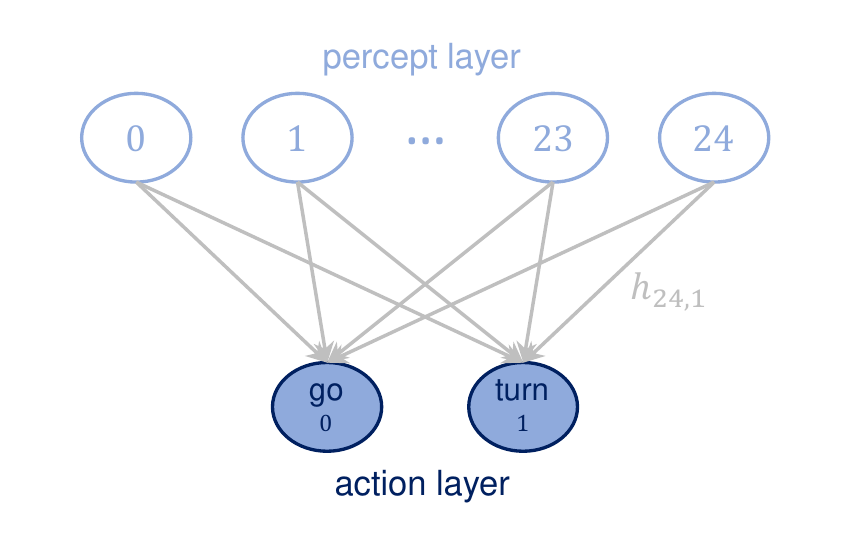}
\caption{Structure of the ECM that consists of two layers, one for the percepts and one for the actions. Percepts and actions are connected by edges whose weight $h_{ij}$ determines the transition probability from the given percept to each action (see Sec.~\ref{sec details of the model} for details on the model).}\label{2layer ps}
\end{figure}

In the PS model, the agent processes the percepts by means of an internal structure called episodic and compositional memory (ECM), whose basic units are called clips and represent an episode of the agent's experience. Mathematically, the ECM can be represented as a directed, weighted graph, where each node corresponds to a clip and each edge corresponds to a transition between two clips. All the edge weights are stored in the adjacency matrix of the graph, termed $h$ matrix. For the purpose of this work, the most basic two-layered structure is sufficient to model simple agents. Percept-clips are situated in the first layer and are connected to the action-clips, which constitute the second layer (see Fig.~\ref{2layer ps}). Let us define these components of the ECM more formally.

\begin{itemize}
\item The \textit{percepts} are mathematically defined as $N$-tuples $s=(s_{1},s_{2},...,s_{N})\in\mathcal{S}$, where $\mathcal{S}$ is the Cartesian product $\mathcal{S}\equiv\mathcal{S}_{1}\times\mathcal{S}_{2}\times...\times\mathcal{S}_{N}$. As it can be seen from this mathematical definition, the percept $s$ has several categories, represented by $\mathcal{S}_{i}$. Each component of the tuple is denoted by $s_{i}\in\{1,...,|\mathcal{S}_{i}|\}$, where $|\mathcal{S}_{i}|$ is the number of possible states of $\mathcal{S}_{i}$. The total number of percepts is thus given by $|\mathcal{S}_{1}|\cdot\cdot\cdot|\mathcal{S}_{N}|$.

\item Analogously, the \textit{actions} are defined as $a=(a_{1},a_{2},...,a_{N})\in\mathcal{A}$, where $\mathcal{A}\equiv\mathcal{A}_{1}\times\mathcal{A}_{2}\times...\times\mathcal{A}_{N}$ and $a_{i}\in\{1,...,|\mathcal{A}_{i}|\}$, where $|\mathcal{A}_{i}|$ is the number of possible states of $\mathcal{A}_{i}$. The total number of actions is given by $|\mathcal{A}_{1}|\cdot\cdot\cdot|\mathcal{A}_{N}|$.
\end{itemize}

As an example, consider an agent that perceives both its internal state, denoted by $\mathcal{S}_{1}$, with two possible percepts $\mathcal{S}_{1}=\{\text{hungry, not hungry}\}$, and some visual input, denoted by $\mathcal{S}_{2}$, with $\mathcal{S}_{2}=\{\text{I see food, I do not see food}\}$. Thus, one out of the four possible percepts could be $s=(\text{hungry, I see food})$. In this case, the possible actions may be $\mathcal{A}=\{\text{go for food, turn around}\}$.

Figure~\ref{2layer ps} represents the structure of the ECM in our model, which consists of a total of 25 percepts and 2 actions (see Sec.~\ref{sec details of the model} for a detailed description).

Let us introduce how the agent interacts with the environment and makes decisions via the ECM. When the agent receives a percept, the corresponding percept-clip inside the ECM is activated, starting a random walk that only ends when an action-clip is reached, which triggers a real action on the environment. The transition probability $P(j|i)$ from a given percept-clip $i$ to an action-clip $j$ is determined by the corresponding edge weight $h_{ij}$ as,
\begin{equation}
P(j|i)=\frac{h_{ij}}{\sum_{k}h_{ik}},\label{prob from hvalues}
\end{equation}
where the normalization is done over all possible edges connected to clip $i$. This process, starting with the presentation of a perceptual input that activates a percept clip and finishing when the agent performs an action on the environment, is termed an (individual) \textit{interaction round}.

The structure of the ECM allows one to easily model learning by just updating the $h$ matrix at the end of each interaction round. Specifically, reinforcement learning is implemented by the environment giving a reward to the agent every time that it performs the correct action. The reward increases the $h$-values\footnote{The $h$ matrix is initialized with all its elements being 1, so that the probability distribution of the actions is uniform for each percept.}, and thus the transition probabilities, of the successful percept-action pair. Hence, whenever the agent perceives again the same percept, it is more likely to reach the correct action. However, in the context of this work, we are setting a learning task in which the agent should perform a sequence of several actions to reach the goal and get the reward. If the reward is given only at the last interaction round, only the last percept-action pair would be rewarded. Thus, some additional mechanism is necessary in order to store a sequence of several percept-action pairs in the agent's memory. This mechanism is called \textit{glow} and the matrix that stores the information about this sequence is denoted by $g$. The components $g_{ij}$, corresponding to the percept-action transition $i\rightarrow j$, are initialized to zero and are updated at the end of \textit{every} interaction round according to:
\begin{equation}
g_{ij}^{(t+1)}=(1-\eta)g_{ij}^{(t)}+\left \{ \begin{matrix} 0 & \mbox{if edge was not traversed}
\\ 1 & \mbox{if edge was traversed,}\end{matrix} \right \}
\end{equation}
where $0\leq \eta\leq 1$ is the glow parameter, which damps the intensity of the given percept-action  memory. For $\eta$ close to one, the actions that are taken at interaction rounds in temporal vicinity to the rewarded action are more intensely remembered that the initial actions. If $\eta=0$, all actions the agent performed until the rewarded interaction are equally remembered. The $g$ matrix is updated in such a way that the percept-action pairs that are used more often to get to the reward are proportionally more rewarded than the pairs that were rarely used. Note that the agent is not able to distinguish an \textit{ordered} sequence of actions, but this is not necessary for the purpose of this work.

In the context of our learning task, the agent receives a reward from the environment at the end of the interaction round at which it reaches a goal. Then, the learning is implemented by updating the $h$ matrix with the rule,
\begin{equation}
h^{(t+1)}=h^{(t)}+R \cdot g,
\end{equation}
where $R \geq 0 $ is the reward (only non-zero if the agent reached the goal at the given interaction round) and $g$ is the updated glow matrix \footnote{Technically, the glow matrix is updated first, and then, if the agent is rewarded, the $h$ matrix is updated.}.

Since we model collective behavior, we consider a group of several agents, each of which has its own and independent ECM to process the surrounding information. Details on the specific learning task and the features of the agents are given in the following section.

\subsection{Details of the model}\label{sec details of the model}
We consider an ensemble of $N$ individuals that we model as PS learning agents, which possess the internal structure (ECM) and the learning capabilities described in section~\ref{PS}. This description of the agents can be seen as a simplified model for species with low cognitive capacities and simple deliberation mechanisms, or just as a theoretical approach to study the optimal behavior that emerges under certain conditions.

With respect to the learning, we set up a concrete task and study the strategy agents develop to fulfill it. In particular, we consider a one-dimensional circular world with sparse resources, which mimics patchy landscapes such as deserts, where organisms need to travel long distances to find food. Inspired by this type of environments, we model a task where agents need to reach a remote food source to get rewarded. The strategy the agents learn via the reinforcement learning mechanism does not necessarily imply that the individual organisms should be able to \textit{learn} to develop it, but can also be interpreted as the optimal behavior that a species would exhibit under the given evolutionary pressures. 

Let us proceed to detail the agents' motor and sensory abilities. The positions that the agents can occupy in the world are discretized $\{0,1,2...W\}$, where $W$ is the world size (total number of positions). Several agents can occupy the same position. At each interaction round, the agent can decide between two actions: either it continues moving in the same direction or it turns around and moves in the opposite direction. The agents move at a fixed speed of 1 position per interaction round. For the remainder of this work, we consider the distance between two consecutive positions of the world to be our basic unit of length. Therefore, unless stated otherwise, all distances given in the following are measured in terms of this unit. We remark that, in contrast to other approaches where the actions are defined with respect to other individuals\footnote{For instance, in the self-propelled particle models \citep{Czirok99,Vicsek95}, the particle changes its orientation at each time step to align itself to the average orientation of the neighboring particles.}, the actions our agents can perform are purely motor and only depend on the previous orientation of the agent. 

\begin{figure}[ht!]
\includegraphics[width=3.4in]{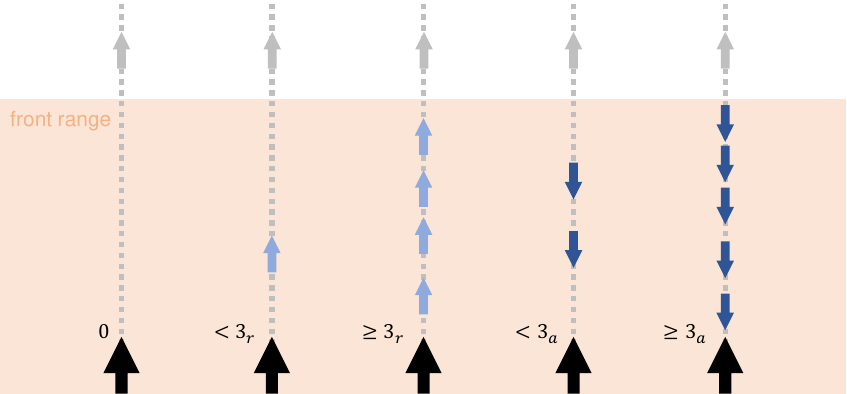}
\caption{Graphical representation of the percepts' meaning. Only the front visual range (colored region) is considered, which corresponds to the values that category $s_{f}$ can take. The focal agent is represented with a larger arrow than the frontal neighbors. The agent can only see its neighbors inside the visual range and it can distinguish if the majority are receding (light blue) or approaching (dark blue) and if they are less or more than three.}\label{drawing percepts}
\end{figure}

Perception is structured as follows: a given agent, termed the focal agent, perceives the relative positions and orientations of other agents inside its visual range\footnote{Radius with center at the agent's position.} $V_{R}$, termed its neighbors. The percept space $S$ (see Sec.~\ref{PS}) is structured in the Cartesian product form $S\in S_{f}\times S_{b}$, where $S_{f}$ is the region in front of the focal agent and $S_{b}$ the region at the back. More precisely, each percept $s=(s_{f},s_{b})$ contains the information of the orientation of the neighbors in each region with respect to the focal agent and if the density of individuals in this region is high or low (see Fig.~\ref{drawing percepts}). Each category of percepts can take the values $s_{f},s_{b}\in\{0,<3_r,\geq3_r,<3_a,\geq3_a\}$ (25 percepts in total), which mean:

\begin{itemize}
\item 0. No agents

\item $<3_r$. There are less than 3 neighbors in this region and the majority of them are receding from the focal agent.

\item $\geq3_r$. There are 3 or more neighbors in this region and the majority of them are receding from the focal agent.

\item $<3_a$. There are less than 3 neighbors in this region and the majority of them are approaching the focal agent.

\item $\geq3_a$. There are 3 or more neighbors in this region and the majority of them are approaching the focal agent.
\end{itemize}

In the following discussions, we refer to the situation where the focal agent has the same orientation as the neighbors as a percept of \textit{positive flow} (majority of neighbors are receding at the front and approaching at the back). If the focal agent is oriented against its neighbors (these are approaching at the front and receding at the back), we denote it as a percept of \textit{negative flow}. Note that the agents can only perceive information about the neighboring agents inside their visual range, but they are not able to see any resource or landmark present in the surroundings. This situation can be found in realistic, natural environments where the distance between resources is large and the searcher has no additional input while moving from one patch to another. Furthermore, the important issue of body orientation is thereby taken into account in our model \citep{Pyke15}.

The interactions between agents are assumed to be sequential, in the sense that one agent at a time receives a percept, deliberates and then takes its action before another agent is given its percept\footnote{In order to do so, agents are given a label at the beginning of the simulation to keep track of the interaction sequence but they are placed at random positions in the world.}. There are two reasons for this choice. For one, in a group of real animals (or other entities), different individuals typically take action at slightly different times, with perfect synchronization being a remarkable and costly exception. The second argument in favor of sequential updating is that it ensures that a given agent's circumstances do not change from the time it receives its percept until the time when its acts. If the actions of all agents were applied simultaneously, a given focal agent would not be able to react on the actions of the other agents in the same round. Such a simplification would not allow us to take into account any sequential, time-resolved interactions between different agents of a group. In the real situation, while one focal agent is deliberating, other agents' actions may change its perceptual input. Therefore, an action that may have been appropriate at the beginning of the round, would no longer be appropriate at this agent's turn.

\begin{figure}[ht!]
\includegraphics[width=3in]{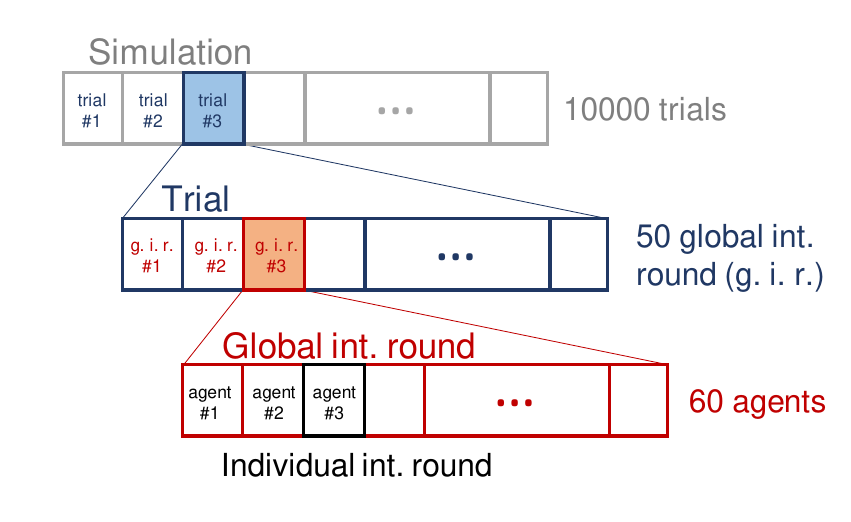}
\caption{Structure of the simulation. Each ensemble of agents is trained for $10^4$ trials, where each trial consists of 50 global interaction rounds (g.i.r.). At each g.i.r., the agents interact sequentially (see text for details).}\label{simstructure}
\end{figure}

\begin{figure}[ht!]
\includegraphics[width=3in]{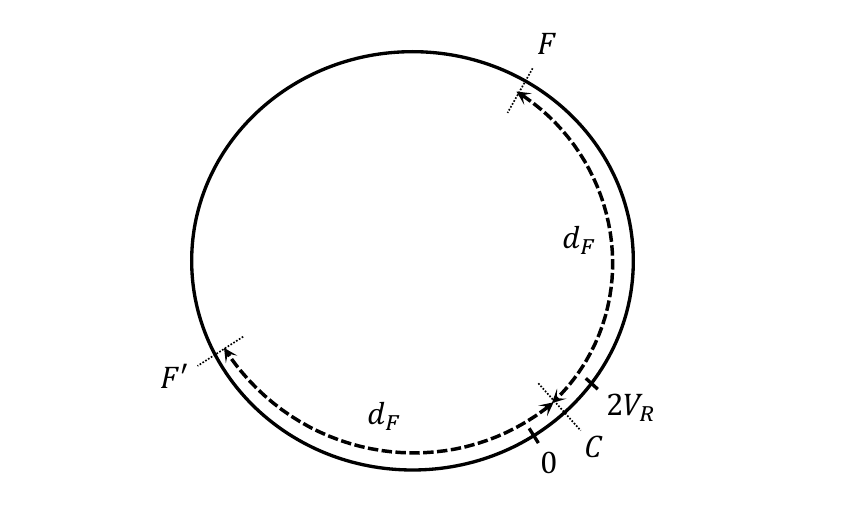}
\caption{1D environment (world). Agents are initialized randomly within the first $2V_{R}$ positions. Food is located at positions $F$ and $F'$. $d_F$ is the distance from the center of the initial region $C$ to the food positions.}\label{1d envir}
\end{figure}

The complete simulation has the structure displayed in Fig.~\ref{simstructure}, where:

\begin{itemize}

\item With each ensemble of $N=60$ agents, we perform a simulation of $10^4$ trials during which the agents develop new behaviors to get the reward (RL mechanism). This process is denoted as \textit{learning process} or \textit{training} from this point on.

\item Each trial consists of $n=50$ global interaction rounds. At the beginning of each trial, all agents of the ensemble are placed in random positions within the initial region (see Fig.~\ref{1d envir}). 

\item We define a global interaction round to be the sequential interaction of the ensemble, where agents take turns to perform their individual interaction round (perception-deliberation-action). Note that each agent perceives, decides and moves only once per global interaction round. 

\end{itemize}

The learning task is defined as follows: at the beginning of each trial, all the agents are placed at random positions within the first $2V_{R}$ positions of the world, with orientations also randomized. Each agent has a fixed number $n$ of interaction rounds over the course of a trial to get to a food source, located at positions $F$ and $F'$ (Fig.~\ref{1d envir}). At each interaction round, the agent first evaluates its surroundings and gets the corresponding percept. Given the percept, it decides to perform one out of the two actions ("go" or "turn"). After a decision is made, it moves one position. If the final position of the agent at the end of an interaction round is a food point, the agent is rewarded ($R=1$) and its ECM is updated according to the rules specified in Sec.~\ref{PS}. Each agent can only be rewarded once per trial.

We consider different learning scenarios by changing the distances $d_{F}$ at which food is positioned. However, note that a circular one-dimensional world admits a trivial strategy for reaching the food without any interactions, namely going straight in one direction until food is reached. Thus, in order to emulate the complexity that a more realistic two-dimensional scenario has in terms of degrees of freedom of the movement, we introduce a noise element that randomizes the orientation of each agent every $s_{r}$ steps\footnote{Not all agents randomize the orientations at the same interaction round, which would lead to random global behavior.} (it changes orientation with probability 1/2). This randomization can be also interpreted biologically as a fidgeting behavior or even as a built-in behavior to escape predators\footnote{Protean movement has been observed in several species \citep{Bile05,Eifler14,Yager90,Combes12} and there exist empirical studies that show that unpredictable turns \citep{Jones11} and complex movement patterns \citep{Richardson18} decrease the risk of predation.} \citep{Humphries70}. If the memory of the organism is not very powerful, we can also consider that, at these randomization points, it forgets its previous trajectory and needs to rely on the neighbors' orientations in order to stabilize its trajectory. The agent can do so, since the randomization takes place right before the agent starts the interaction round.

Under these conditions, we study how the agents get to the food when the only input information available to them is the orientation of the agents around them.

\section{Learned behavior in different scenarios}\label{learning diff scen}

We consider different learning scenarios characterized by the distance $d_{F}$ (see Fig.~\ref{1d envir}). We study how the dynamics that the agents develop in order to reach the food source change as the distance $d_F$ increases. In particular, we focus on two extreme scenarios: one where the resource is within the initial region ($d_F<V_R$) ---agents are initialized within the first $2V_R$ positions of the world---, and the other one where the resource is at a much larger distance. As a scale for this distance, we consider how far an agent can travel on average with a random walk, which is $d_{rw}=\sqrt{n}$ providing that it moves one position per interaction round. Hence, the other extreme scenario is such that $d_F \gg d_{rw}$. The first situation, where $d_F<V_R$, mimics an environment with densely distributed resources, whereas the second one ($d_F \gg d_{rw}$) resembles a resource-scarce environment where a random walk is no longer a valid strategy for reaching food sources. 

\begin{table}[htb]
\begin{tabular}{|l|c|l|c|}
\hline
\multicolumn{2}{|c|}{Agent} & \multicolumn{2}{|c|}{Environment}\\ \hline
\multicolumn{1}{|c|} {Description} & Value &\multicolumn{1}{|c|} {Description}& Value \\ \hline 
Visual range ($V_R$) & 6 & Number of agents ($N$) & 60 \\ \hline
Reorient. freq. ($s_r$)  & 5 & World size ($W$) & 500 \\ \hline
Glow ($\eta$) & 0.2 & Int. rounds per trial ($n$) & 50 \\ \hline
Reward ($R$) & 1 & Number of trials & $10^4$ \\ \hline
\end{tabular}
\caption{Description of the parameters used in the learning simulations with PS.}
\label{tabla:parameters}
\end{table}

\begin{figure}[ht!]
\includegraphics[width=3.4in]{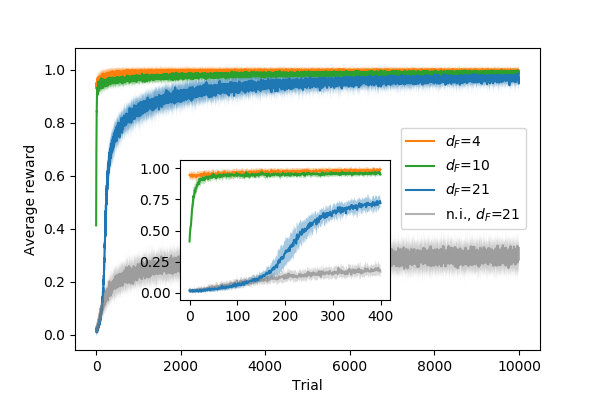}
\caption{Learning curves for $d_{F}=4,10,21$ and $d_F=21$ for non-interacting (n.i.) agents. The curve shows the percentage of agents that reach the food source and obtain a reward of $R=1$ at each trial. For each task, the average is taken over 20 (independent) ensembles of 60 agents each and the shaded area indicates the standard deviation. Zooming into the initial phase of the learning process, the inset figure shows a faster learning in the task with $d_F=10$ than in the task with $d_F=21$. In the case of $d_F=21$, no agent is able to reach the food source in the first trial, and it takes the interacting agents approx. 200 trials to outperform the n.i. agents.}\label{Learning-curves}
\end{figure}

The parameters of the model that are used in all the learning processes are given in Table~\ref{tabla:parameters}. Providing that $d_{rw}=\sqrt{50}\simeq 7$, we consider values of $d_{F}$ ranging from 2 to 21 and focus on the cases with $d_F=4,21$ as the representative examples of resource-dense and resource-scarce environments, respectively. All agents start the learning process with a newly initialized $h$ matrix, so they perform each action ("go" or "turn") with equal probability. Figure~\ref{Learning-curves} shows the learning curves for three different scenarios, where the food is placed at $d_F=4,10,21$. The learning processes are independent from each other, that is, the distance $d_F$ does not change within one complete simulation of $10^4$ trials. In this way, we can analyze the learned behaviors separately for each $d_F$. The learning curve displays the percentage of agents that reach the food source and obtain a reward at each trial. As a baseline for comparison, we also set the same learning task with $d_F=21$ for non-interacting (n.i.) agents (we set $V_{R}=0$, so they cannot see the neighbors). The n.i. agents learn to go straight almost deterministically\footnote{Therefore, the agent performs a random walk with $n/s_r=50/5=10$ steps of length $s_r=5$, which allows it to cover a distance of $5\sqrt{10}\simeq16$ positions.} ---the probability for the action "go" at the end of the learning process is almost $1$ for percept $(0,0)$---. The rest of percepts are never encountered, so the initial $h$ values remain the same. Due to the periodic randomization of the agents' orientation, it can be seen that they do not reach the efficiency rate of the interacting agents (see figure~\ref{Learning-curves}) and only one out of three agents reaches the reward at each trial. Figure~\ref{Learning-curves} shows that, for $d_F=4$, the food source is so close (inside the initial region) that the agents get the reward in all the trials from the beginning. On the other hand, the tasks with $d_F=10,21$ show a learning process that takes more trials for the agents to come up with a behavior that allows them to get to the reward. In particular, only $40\%$ of the agents are able to reach the goal with the initial behavior (Brownian motion) in the scenario with $d_F=10$ and this percentage drops to almost $0\%$ in the case with $d_F=21$. Note that it takes more trials for the agents to learn how to get to the furthest point ($d_F=21$) than it takes for $d_F=10$ (see inset in Fig.~\ref{Learning-curves}). The interacting agents start outperforming the n.i. agents in the task with $d_F=21$ at trial 200, where they start to form aligned swarms, as one can also see from the increase in the alignment parameter at the same trial in Fig.~\ref{evol align} (see Sec.~\ref{subsec alignment} for more details).

\subsection{Individual responses}\label{SEC ind responses}

\begin{figure}[ht!]
\centering
\subfigure[$d_{F}=21$]{\includegraphics[width=3.4in]{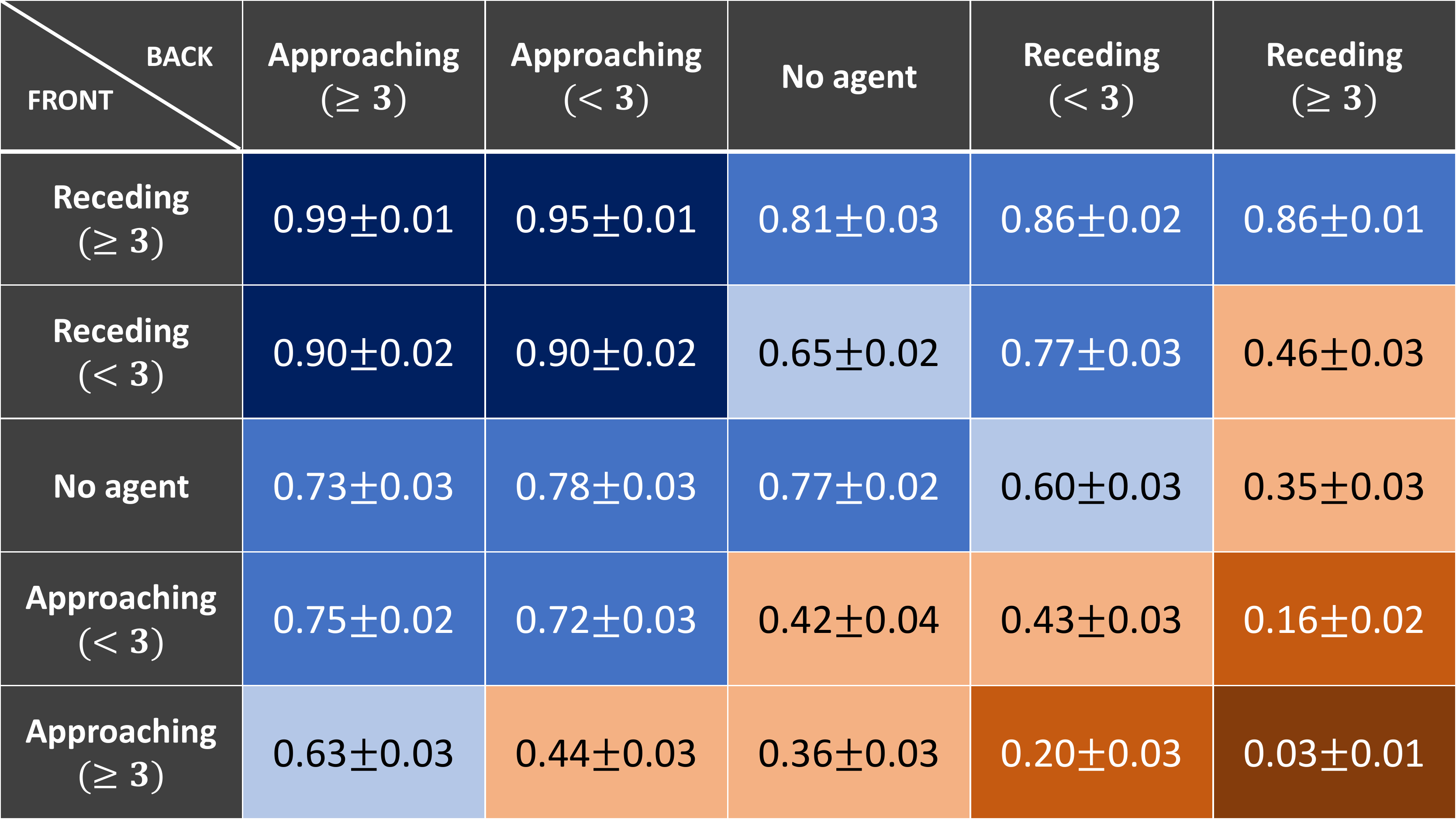}}
\subfigure[$d_{F}=4$]{\includegraphics[width=3.4in]{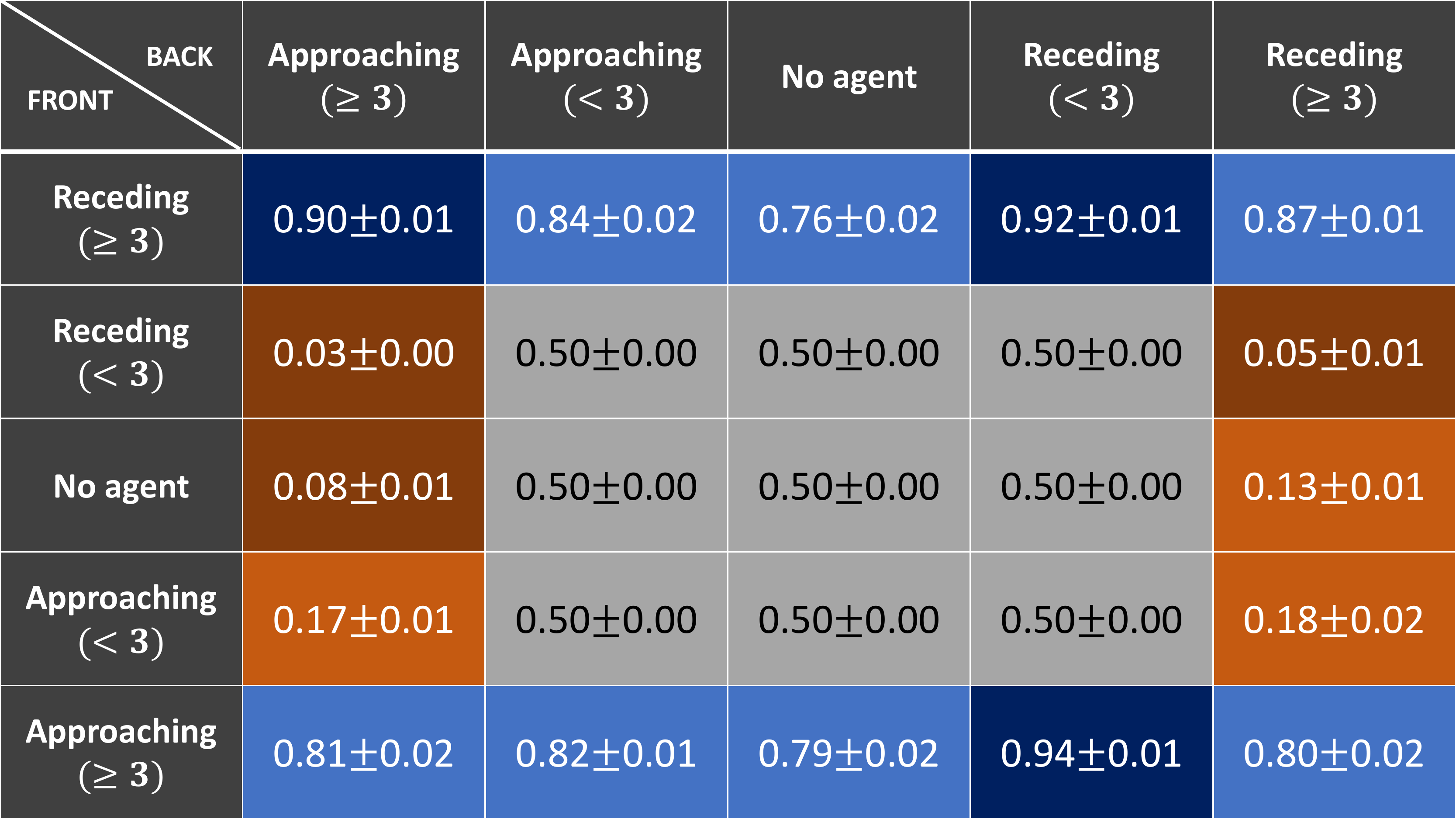}}
\subfigure[Ideal probabilities for alignment]{\includegraphics[width=1.2 in]{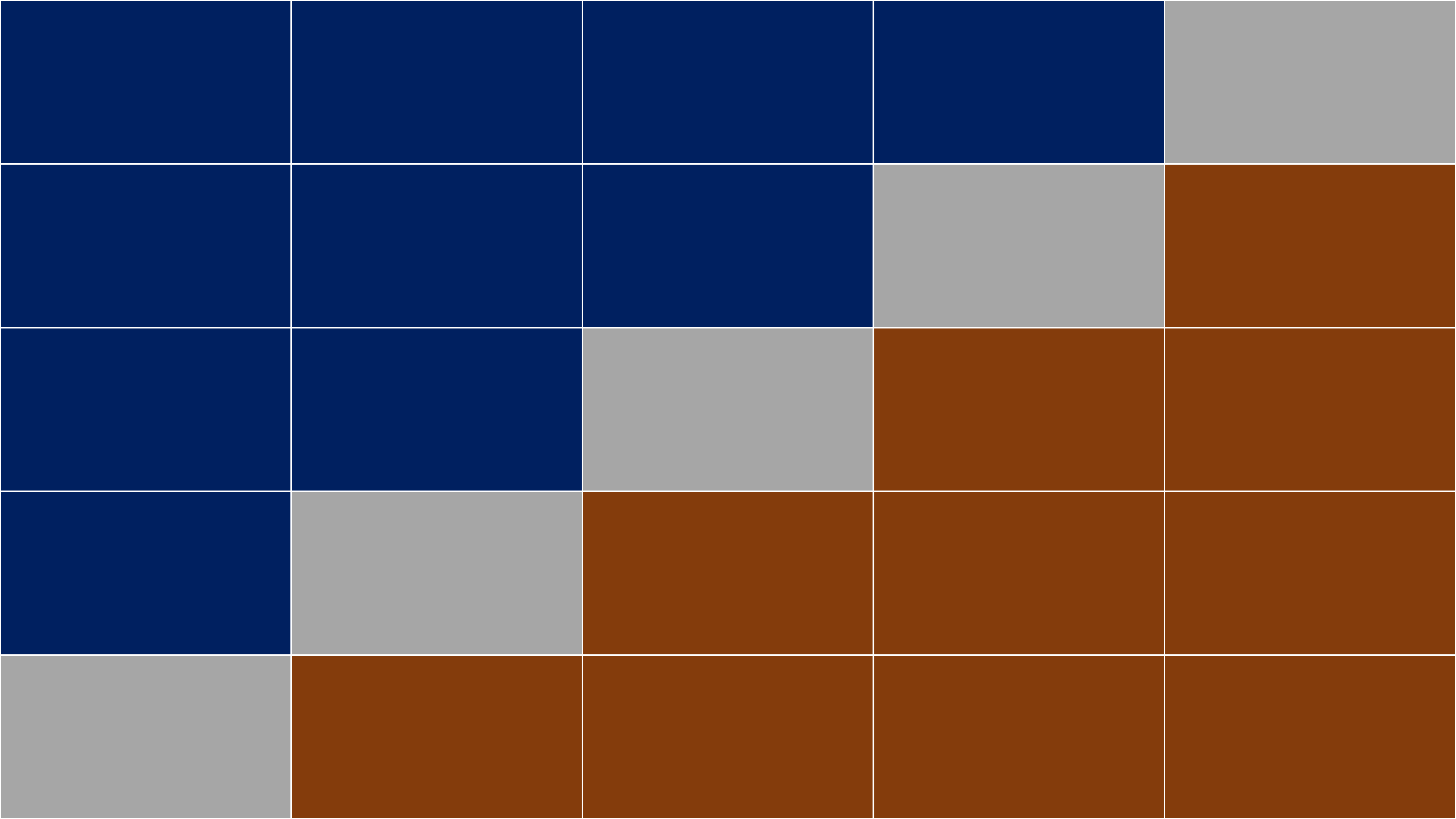}}\hspace{0.5 in}
\subfigure[Ideal probabilities for cohesion]{\includegraphics[width=1.2 in]{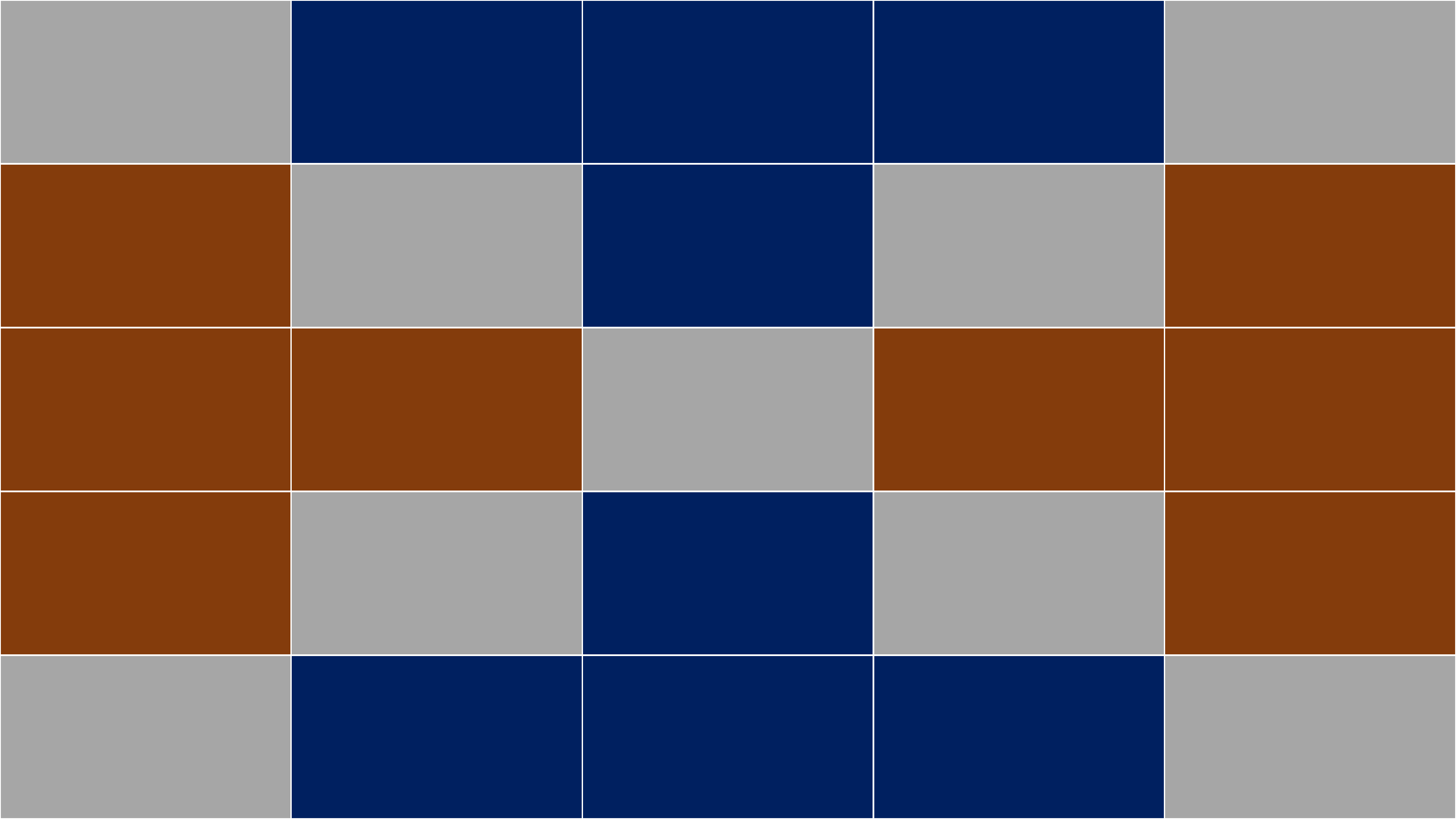}}
\caption{Learned behavior at the end of the training process. The final probabilities in the agents' ECM for the action "go" are shown for each of the 25 percepts ($5x5$ table). Tables (a) and (b) show the final probabilities learned in the scenarios with $d_{F}=21$ and $d_F=4$ respectively. The average is taken over 20 ensembles (each learning task) of 60 agents each. Background colors are given to easily identify the learned behavior, where blue denotes that the preferred action for that percept is "go" and orange denotes that it is "turn". More specifically, the darker the color is, the higher the probability for that action, ranging from grey ($p \simeq 0.5$), light ($0.5<p<0.7$) and normal ($0.7\leq p<0.9$) to dark ($p\geq 0.9$). Figures (c) and (d) show what the tables would look like if the behavior is purely based on alignment (agent aligns to its neighbors with probability 1) or cohesion (agent goes towards the region with higher density of neighbors with probability 1), respectively. See text for details.}\label{Learned-probabilities}
\end{figure}

The behavior the agents have learned at the \textit{end} of the training can be studied by analyzing the final state of the agents' ECM, from where one obtains the final probabilities for each action depending on the percept the agents get from the environment (see Eq.~\eqref{prob from hvalues}). These final probabilities are given in Fig.~\ref{Learned-probabilities} for the learning tasks with $d_F=4,21$. 

Tables of figure~\ref{Learned-probabilities} show the probability of taking the action "go" for each of the $25$ percepts. We focus on the learning tasks with $d_F=4,21$, which represent the two most distinctive behaviors that we observe.

Let us start with the case of $d_F=21$ (Fig.~\ref{Learned-probabilities} (a)), which corresponds to a task where the food is located much further away than the distance reachable with a random walk. In this case, highly aligned swarms emerge as the optimal collective strategy for reaching the food (see also figs.~\ref{evol align} and \ref{Trajectories}), since the orientations of the surrounding neighbors allow the focal agent to stabilize its orientation against the periodic randomization. The individual responses that lead to such collective behavior can be studied by looking at table (a): the diagonal corresponds to percepts with a clear reaction leading to alignment, i.e. to keep going when there is a positive flow of neighbors and to turn if there is a negative flow. More specifically, one can see that when the agent is in the middle of a swarm and aligned with it, the probability that it keeps going is 0.99 for dense swarms [percept ($\geq 3_r,\geq 3_a$)] and 0.90 for sparse swarms [percept ($<3_r,<3_a$)]. In the same situations, the agent that is not aligned turns around with probability 0.97 for dense swarms [percept ($\geq 3_a,\geq 3_r$)] and 0.57 for sparse swarms [percept ($<3_a,<3_r$)]. Outside the diagonal, one observes that the probability of turning is high when a high density of agents are approaching the focal individual from the front (last row) and the agents in the back are not approaching. We can also analyze the learned behavior at the back edge of the swarm, which is important to keep the cohesion of the swarm. When an agent is at the back of a dense swarm and aligned with it [percept ($\geq 3_r,0$)], the probability of keeping the orientation is 0.81. If instead, the agent is oriented against the swarm [percept ($0,\geq 3_r$)] the probability of turning around to follow the swarm is 0.65. This behavior is less pronounced when the swarm is not so dense [percepts ($<3_r,0$), ($0,<3_r$)], in fact, when a low density of neighbors at the back are receding from the focal agent [percept ($0,<3_r$)], the focal agent turns around to rejoin the swarm with probability 0.4, which results in this agent leaving the swarm with higher probability. If the agent is alone [percept ($0,0$)], it keeps going with probability 0.77.

\begin{figure}[ht!]
\includegraphics[width=3 in]{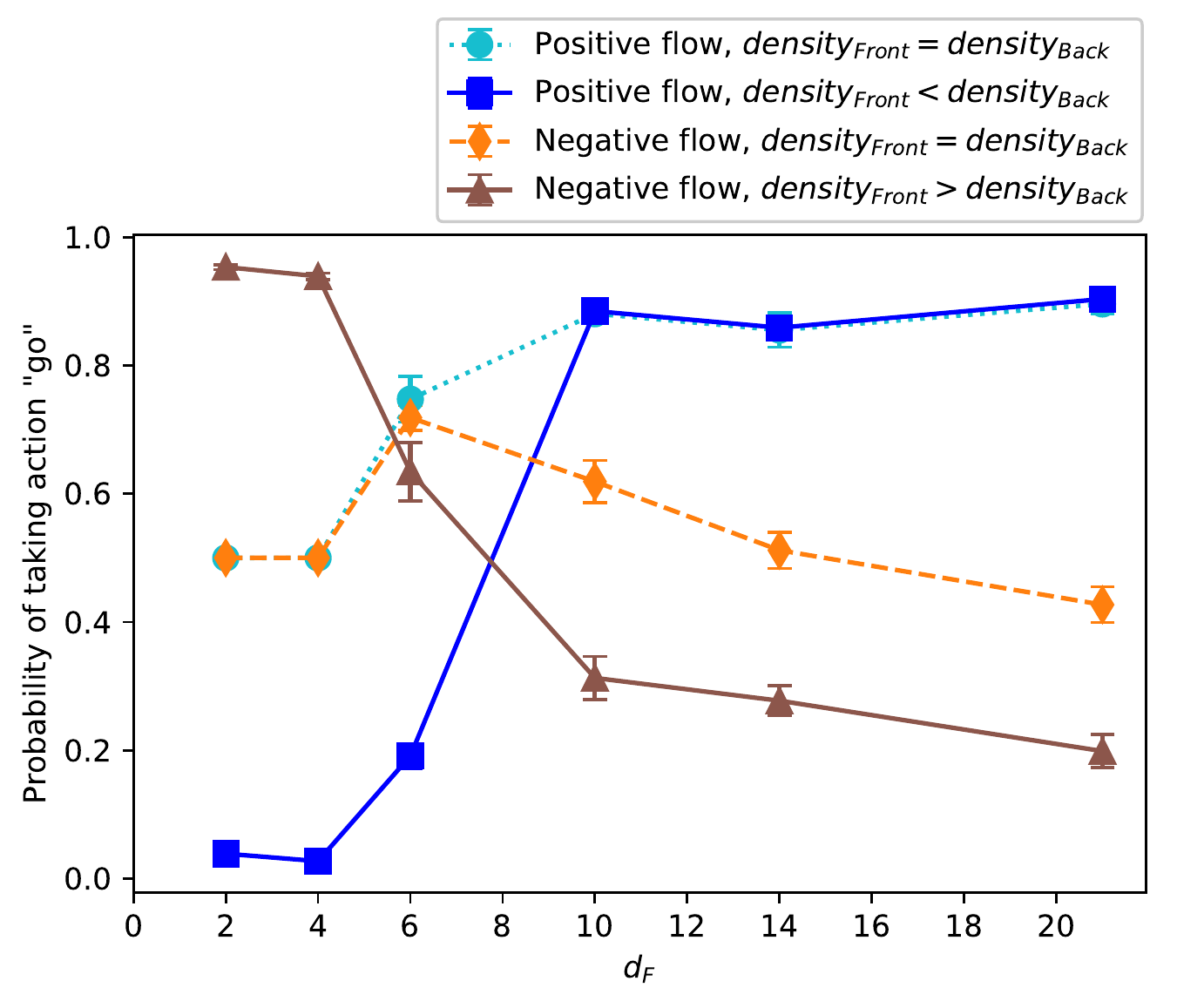}
\caption{Final probability of taking the action "go" depending on the learning task (increasing distance to food source $d_F$) for four significant percepts. The percepts are $(<3_r,<3_a),(<3_r,\geq 3_a),(<3_a,<3_r),(\geq 3_a,<3_r)$, respectively (see legend). The average is taken over the agents' ECM of 20 independently trained ensembles (1200 agents) at the end of the learning process. Each ensemble performs one task per simulation ($d_F$ does not change during the learning process).}\label{comparison_matrixdf}
\end{figure}

A very different table is observed for $d_F=4$ (Fig.~\ref{Learned-probabilities} (b)). In this task, the food source is located inside the initial region where the agents are placed at the beginning of the trials, so the agents perceive, in general, high density of neighbors around them. For this reason, they rarely encounter the nine percepts encoding low density ---that correspond to the ones at the center of the table, with grey background (Table (b) in Fig.~\ref{Learned-probabilities})--- throughout the interaction rounds they perform until they get the reward. The corresponding probabilities are the initialized ones, i.e. $1/2$ for each action. For the remaining percepts, we observe that the agents have learned to go to the region with higher density of neighbors, which leads to very cohesive swarms (see also Sec.~\ref{cohesion section}). Since the food source is placed inside the initialization region in this case ---which is also within the region agents can cover with a random walk---, there is a high probability that there are several agents already at the food source when an agent arrives there, so they learn to go to the regions with higher density of agents. This behavior can be observed, for instance, for percepts in the first column (high density at the back) and second, third and fourth row (low/no density at the front), where the agents turn around with high probability. In addition, we observe that there is a general bias towards continuing in the same direction, which can be seen for example in percepts with the same density in both regions (e.g. percepts at the corners of the table). The tendency to keep walking is always beneficial in one-dimensional environments to get to the food source (non-interacting agents learn to do so deterministically, as argued for Fig.~\ref{Learning-curves}). In general, we observe that, in order to find the resource point at $d_F=4$, agents do not need to align with their neighbors because the food is close enough that they can reach it by performing a Brownian walk. 

Figures~\ref{Learned-probabilities} (c) and (d) show what the tables would look like if the agents had deterministically (with probability 1) learned just to align with the neighbors (c) or just to go to the region inside the visual range with higher density of neighbors (d). In these figures, percepts for which there is no pronounced optimal behavior have grey background.

In Fig.~\ref{comparison_matrixdf}, we select four representative percepts that show the main differences in the individual behaviors and plot the average probability of taking the action "go" at the end of a wide range of different learning scenarios where the distance to the food source is increasingly large. We observe that there are two clear regimes with a transition that starts at $d_F=6$. This is the end of the initial region (see Fig.~\ref{1d envir}, with $V_R=6$ in our simulations) where the agents are positioned at the beginning of each trial (see appendix~\ref{APP transition} for details on why this transition occurs at $d_F=6$). The main difference between regimes is that, when the food is placed near the initial positions of the agents, they learn to "join the crowd", whereas, if the food is placed farther away, they learn to align themselves and "go with the flow". More specifically, for $d_F<6$, the orientations of the surrounding neighbors do not play a role, but the agents learn to go to the region (front/back) with higher number of neighbors, which leads to unaligned swarms with high cohesion. On the contrary, for the tasks with $d_F>6$, the agents tend to align with their neighbors.  This difference in behavior can be observed, for instance, in the dark blue (squares) curve of figure~\ref{comparison_matrixdf}, which corresponds to the percept "positive flow and higher density at the back". We observe that for $d_F=2,4$, the preferred action is "turn" (the probability of taking action "go" is low), since there are more neighbors at the back. However, for $d_F=10,14,21$, the agents tend to continue in the same direction, since there is a positive flow (neighbors have the same orientation as the focal agent). Analogously, the brown curve (triangles) shows the case where there is a negative flow and higher density at the front, so agents trained to find nearby food ($d_F=2,4$) have high probability of going, whereas agents trained to find distant food ($d_F=10,14,21$) have high probability of turning.

In general, we observe that agents with the same motor and sensory abilities can develop very different behaviors in response to different reward schemes. Agents start with the same initial ECM in all the learning scenarios, but depending on the environmental circumstances, in our case the distance to food, some responses to sensory input happen to be more beneficial than others in the sense that they eventually lead the agent to get a reward. For instance, agents that happen to align with their neighbors are the ones that reach the reward when the food is far away, so this response is enhanced in that particular scenario, but not in the one with nearby food.

\subsection{Collective dynamics}

\begin{figure}[ht!]
\includegraphics[width=3.4in]{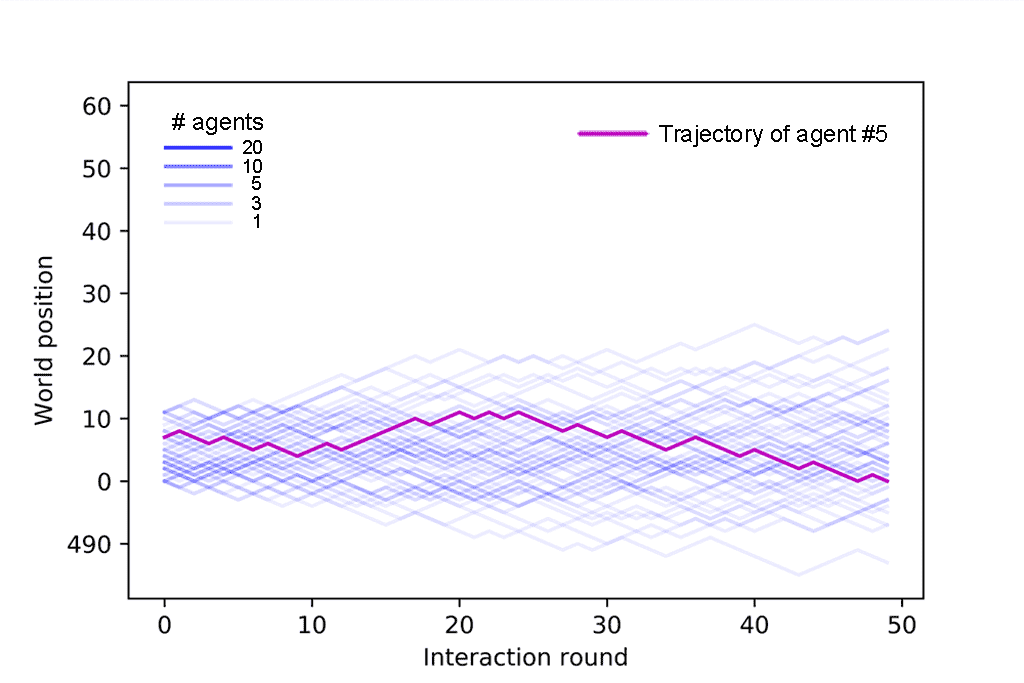}
\caption{Trajectories (position vs. time) of an ensemble of 60 agents in one trial prior to any learning process. The vertical axis displays the position of the agent in the world and the horizontal axis the interaction round (note that the trial consists of $n=50$ rounds). Each line corresponds to the trajectory of one agent. However, some agents' trajectories overlap, which is indicated by the color intensity. The trajectory of one particular agent is highlighted for clarity.}\label{traj first trial}
\end{figure}

\begin{figure}[ht!]
\subfigure[Trajectories after training with $d_F=21$]{\includegraphics[width=3.4in]{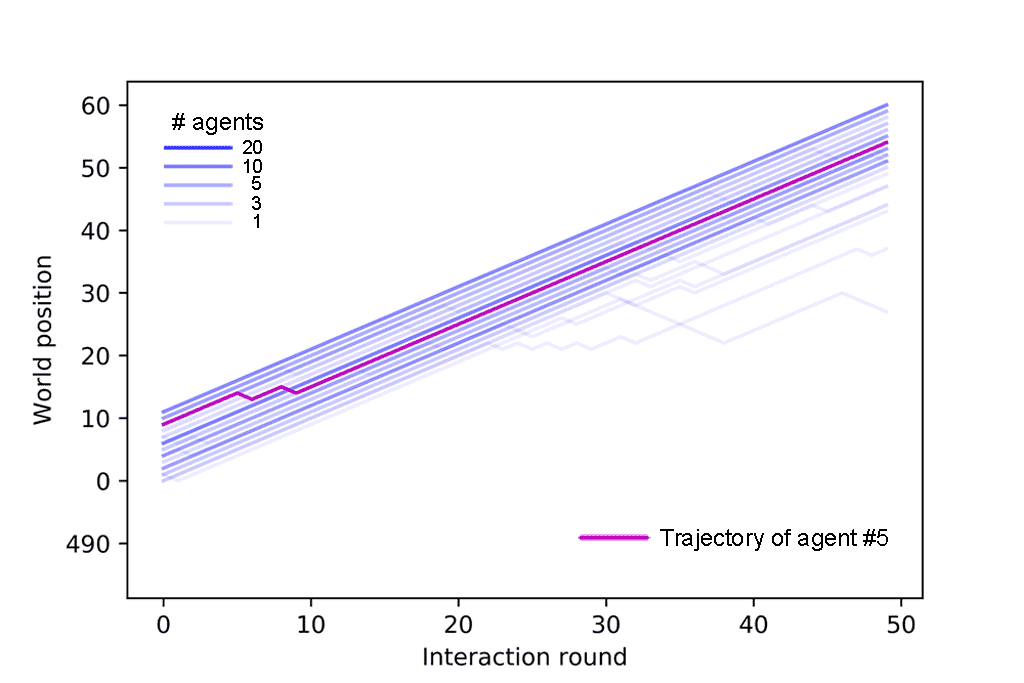}}
\subfigure[Trajectories after training with $d_F=4$]{\includegraphics[width=3.4in]{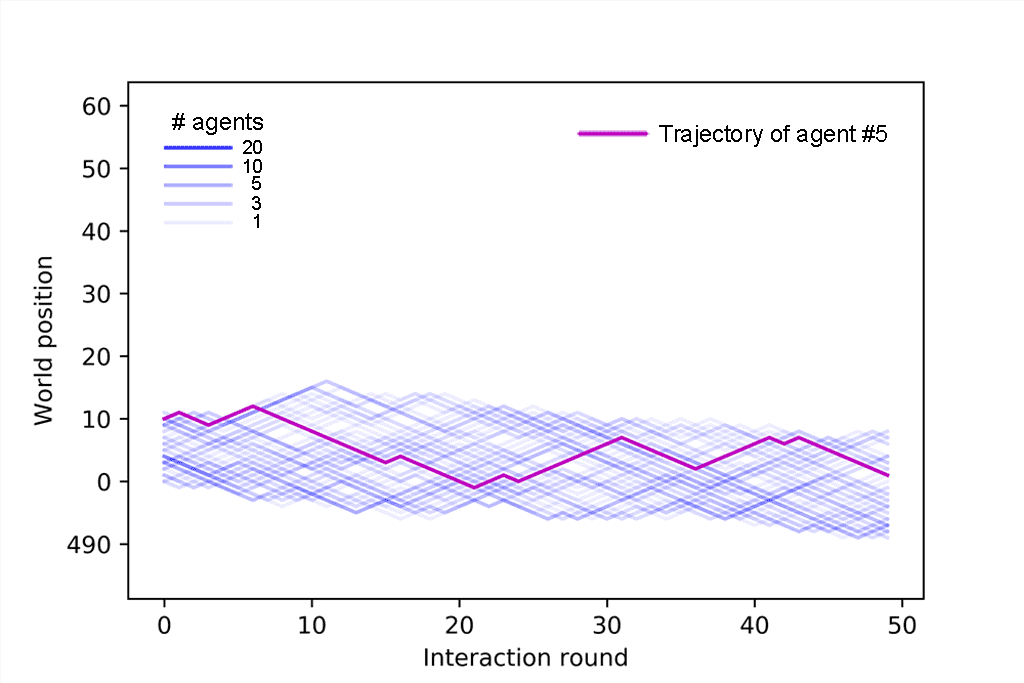}}
\caption{Trajectories of all agents of an ensemble in the last trial of the learning process for (a) $d_{F}=21$ and (b) $d_F=4$. Ensembles of agents trained to find distant food form aligned swarms (a), whereas agents trained to find nearby food form cohesive, unaligned swarms (b). With the same number of interaction rounds, aligned swarms (a) cover larger distances than cohesive swarms (b). In addition, observe that trajectories in panel (b) spread less than in Fig.~\ref{traj first trial}.}\label{Trajectories}
\end{figure}

In this section, we study the properties of the collective motion that emerges from the learned individual responses analyzed in the previous section. We focus on two main properties of the swarms, namely alignment and cohesion. Figures~\ref{traj first trial} and~\ref{Trajectories} show the trajectories of the agents of one ensemble before (Fig.~\ref{traj first trial}) any learning process and at the end of the learning processes with $d_F=4,21$ (Fig.~\ref{Trajectories}). One can see that the collective motion developed in the two scenarios differs greatly in terms of alignment and cohesion. Thus, we quantify and analyze these differences in the following.

\subsubsection{Alignment}\label{subsec alignment}

The emergence of aligned swarms as a strategy for reaching distant resources is studied by analyzing the order parameter, defined as
\begin{equation}
\phi=\frac{1}{N}|\sum_{i=1..N}v_{i}|, \label{eq align par}
\end{equation}
where $N$ is the total number of agents and $v_{i}\in\{1,-1\}$ the orientation of each agent (clockwise or counterclockwise). The order parameter or \textit{global alignment parameter} goes from 0 to 1, where 0 means that the orientations of the agents average out and 1 means that all of them are aligned. In addition, we also evaluate the \textit{local} alignment parameter, since the visual perception of the agent only depends on its local surroundings, and so does the action it takes. In this case, the order parameter $\phi_{i}$ is computed for each agent $i$, considering only the orientation of its neighbors.

\begin{figure}[htb!]
\includegraphics[width=3.4in]{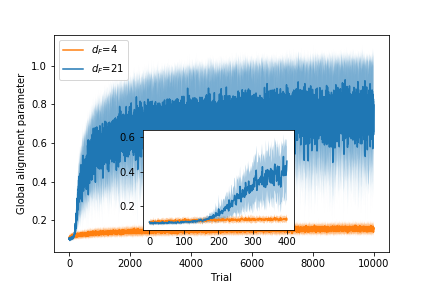}
\caption{Evolution of the global alignment parameter through the learning processes with $d_F=4,21$. At each trial, there is one data point that displays the average of the order parameter, first over all the (global) interaction rounds of the trial and then over 20 different ensembles of agents, where each ensemble learns the task independently. Shaded areas represent one standard deviation.}\label{evol align}
\end{figure}

Figure~\ref{evol align} shows how agents that need to find nearby food do not align, whereas those whose task is to find distant resources learn to form strongly aligned swarms as a strategy for getting the reward, as can be seen from the increase in the order parameter over the course of the training. The inset in Fig.~\ref{evol align} shows that agents with the reward at $d_F=21$ start to align with the neighbors from trial 200, which leads to the conclusion that increasing the alignment is the behavior that allows them to get to the reward (note that the agents start to be rewarded also from trial 200, as can be seen in the inset of Fig.~\ref{Learning-curves}). The large standard deviation in the $d_F=21$ case is due to the fact that, in some trials, agents split in two strongly aligned groups that move in opposite directions (see Fig.~\ref{splitting} (a) in appendix~\ref{APP alignment} for details). 

\subsubsection{Cohesion}\label{cohesion section}
In this section, we study the cohesion and stability of the different types of swarms. In particular, we quantify the cohesion by means of the average number of neighbors (agents within visual range of the focal agent),
\begin{equation}
M=\frac{1}{N}\sum^N_{i=1} m_i,
\end{equation}
where $m_i$ is the number of neighbors of the $i$th agent.

\begin{figure}[htb!]
\includegraphics[width=3.4in]{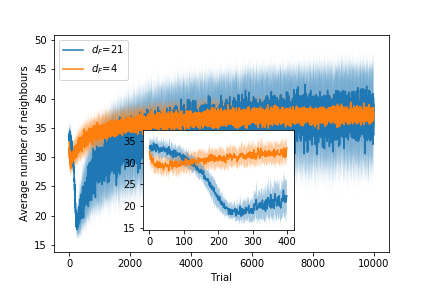}
\caption{Evolution of the average number of neighbors around each agent through the learning processes with $d_F=4,21$. At each trial, there is one data point that displays the average of $M$, first over all the (global) interaction rounds of the trial and then over 20 different ensembles of agents, where each ensemble learns the task independently. Shaded areas represent one standard deviation.}\label{evol cohesion}
\end{figure}

Figure~\ref{evol cohesion} shows the evolution of the average number of neighbors through the learning processes with $d_F=4,21$. In the training with $d_F=21$, we observe a decay in $M$ in the first 200 trials, due to the fact that agents start to learn to align locally (see appendix~\ref{APP cohesion} and Fig.~\ref{app fig local evol} therein for details), but the global alignment is not high enough to entail an increase in the average number of neighbors. Therefore, as agents begin to move in straight lines for longer intervals (instead of the initial Brownian motion), they tend to leave the regions with a higher density of agents and $M$ drops. From trial 200 onwards, agents start to form aligned swarms ---global alignment parameter increases (see inset of Fig.~\ref{evol align})--- to get to the food, which leads to an increase in $M$ (see inset of Fig.~\ref{evol cohesion}). In the training with $d_F=4$, agents learn quickly (first 50 trials) to form cohesive swarms, so $M$ increases until a stable value of 36 neighbors is attained.

Up to this point, all the analyses have been done with trials of $50$ interaction rounds. However, this is insufficient for assessing the stability of the swarm. For this purpose, we take the already trained ensembles and let them walk for longer trials\footnote{The agents do not learn anything new in these simulations, i.e. their ECMs remain unchanged.} so that we can analyze how the cohesion of the different swarms evolves with time. We place the agents (one ensemble of 60 agents per simulation) in a world that is big enough so that they cannot complete one cycle within one trial. This resembles infinite environments insofar as agents that leave the swarm have no possibility of rejoining it. This allows us to study the stability of the swarm cohesion and the conditions under which it disperses.

\begin{figure}[htb!]
\subfigure[Ensemble trained with $d_F=21$]{\includegraphics[width=3.4in]{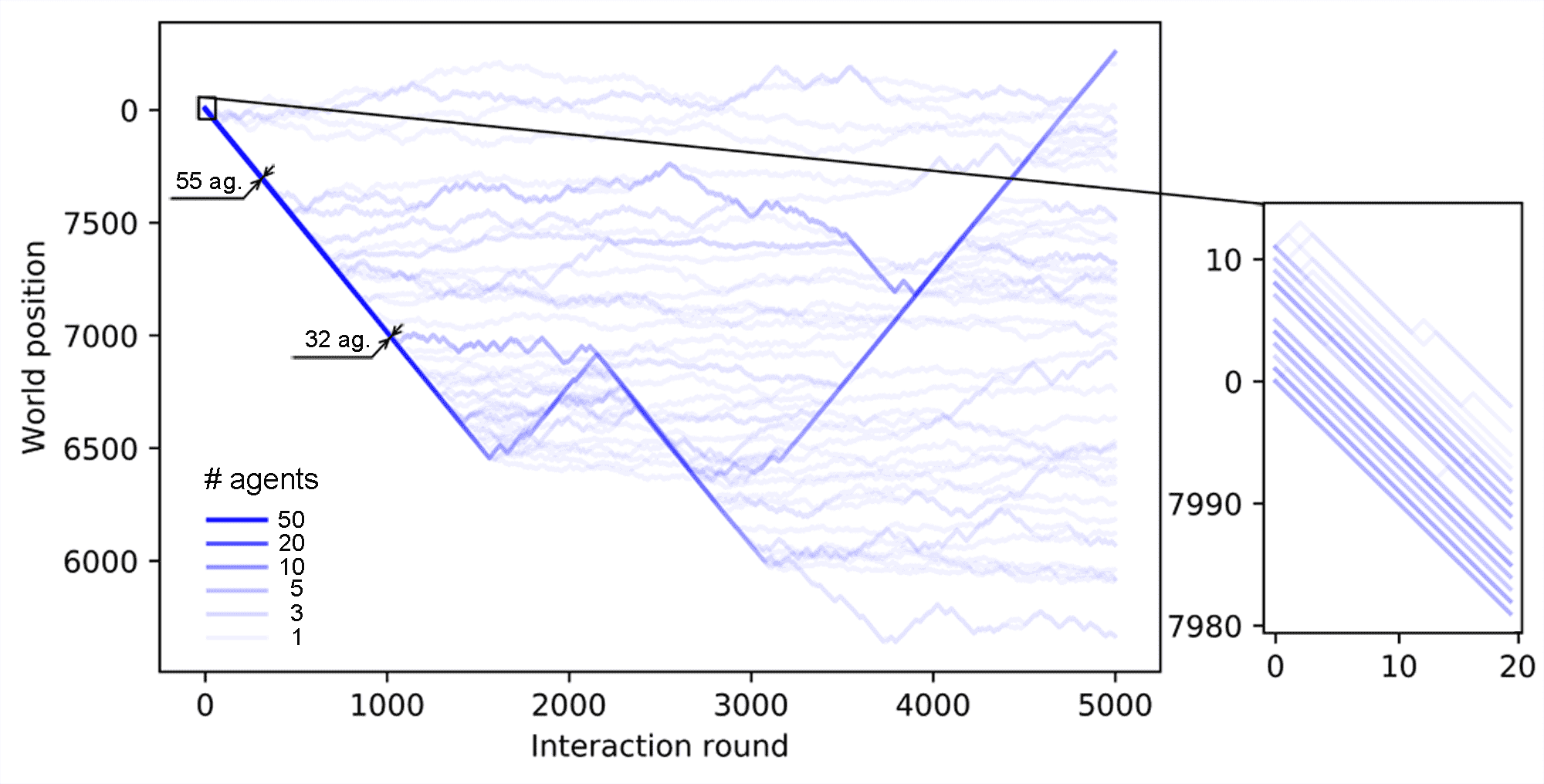}}
\subfigure[Ensemble trained with $d_F=4$]{\includegraphics[width=3.4in]{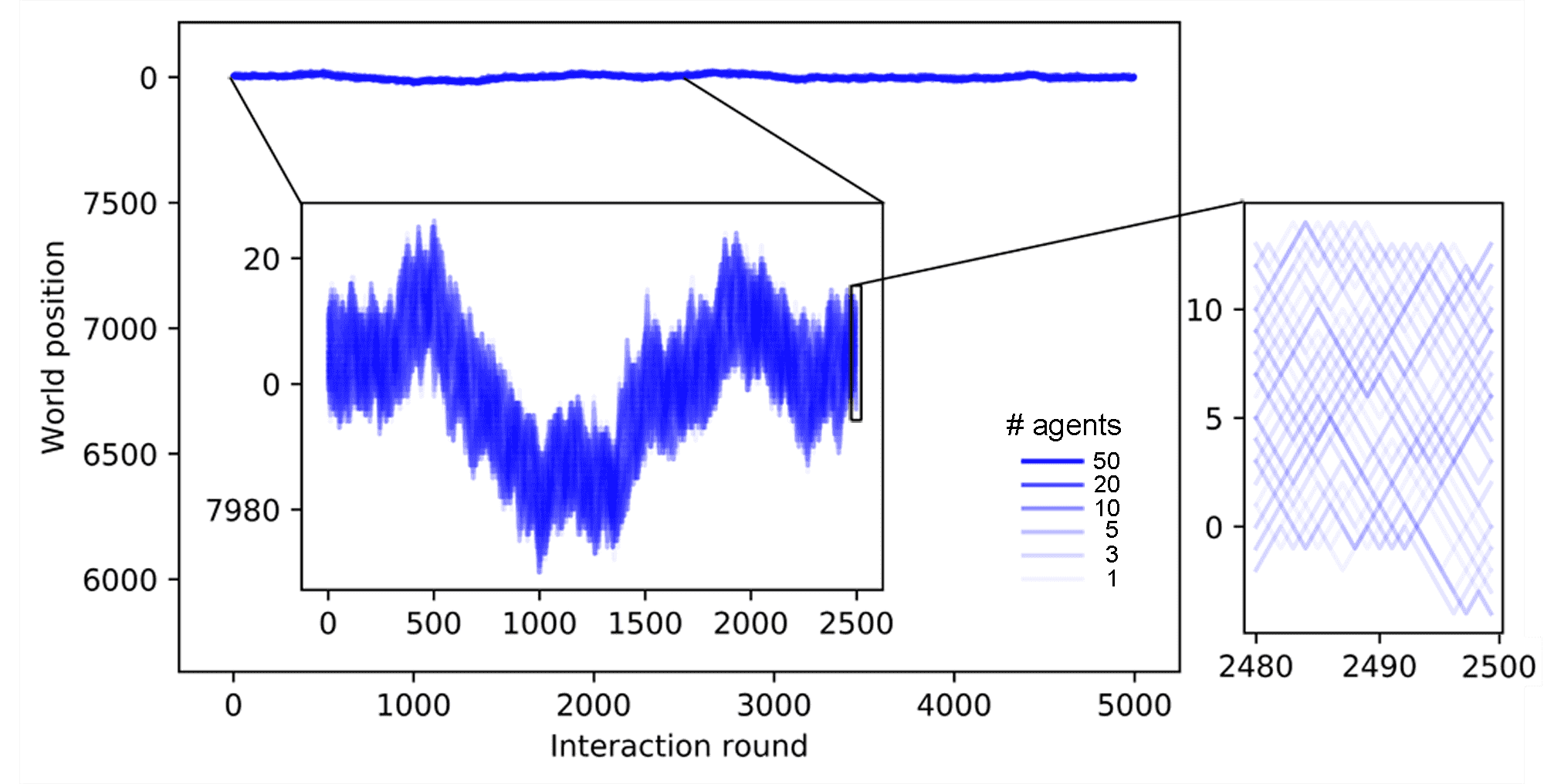}}
\caption{Trajectories of an ensemble of 60 agents, in a world of size $W=8000$, shown over 5000 interaction rounds. (a) Agents trained with $d_F=21$ form a swarm that continuously loses members until it dissolves completely. (b) Agents trained with $d_F=4$ form a highly cohesive swarm for the entire trial. The centered inset of this plot shows the first 2500 rounds, with a re-scaled vertical axis to observe the movement of the swarm. Insets on the right zoom in to 20 interaction rounds so as to resolve individual trajectories.}\label{drop cohesion}
\end{figure}

\begin{figure}[htb!]
\includegraphics[width=3.4in]{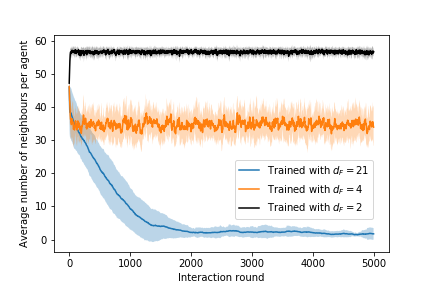}
\caption{Evolution of the average number of neighbors throughout the trial of 5000 interaction rounds (average is taken over 20 ensembles of 60 agents each, where for each ensemble the simulation is performed independently. Shaded areas indicates one standard deviation).}\label{num neigh cohesion}
\end{figure}

Figure~\ref{drop cohesion} shows the trajectories of ensembles of agents trained with different distances $d_F$. In the case with $d_F=21$ (Fig.~\ref{drop cohesion} (a)), there is a continuous drop of agents from the swarm until the swarm completely dissolves. On the other hand, agents trained with $d_F=4$ (Fig.~\ref{drop cohesion} (b)) present higher cohesion and no alignment (see inset of Fig.~\ref{drop cohesion} (b)). Note that this strong cohesion makes individual trajectories spread less than the Brownian motion exhibited by agents prior to the training (see Fig.~\ref{traj first trial}). The evolution of the average number of neighbors throughout the simulation is given in figure~\ref{num neigh cohesion}, where we compare the cohesion of ensembles of agents trained with $d_F=2,4,21$. In the latter case, the agents leave the swarm continuously, so the average number of neighbors decreases slowly until the swarm is completely dissolved. For $d_F=2$ ($d_F=4$) the individual responses are such that the average number of neighbors increases (decreases) in the first 30 rounds until the swarm stabilizes and from then on $M$ stays at a stable value of $57$ ($35$) neighbors. The average number of neighbors is correlated to the swarm size, which we measure by the difference between the maximum and minimum world positions occupied by the agents (modulo world size $W=500$). As one can see in Fig.~\ref{drop cohesion} (b), all agents remain within the swarm. If the swarm size increases, the average number of neighbors decreases, since the agents are distributed over a wider range of positions. The swarm stabilizes at a given size depending on the individual responses learned during the different trainings. For instance, swarms formed by agents trained with $d_F=2$ stabilize at swarm sizes of approximately 9 positions, whereas those trained with $d_F=4$ stabilize at larger swarm sizes (around 17 positions, see e.g. inset of fig.~\ref{drop cohesion} (b)), which explains the lower value of $M$ observed in Fig.~\ref{num neigh cohesion} for $d_F=4$. 

\subsubsection{Comparison between learning scenarios}
\begin{figure}[htb!]
\includegraphics[width=3 in]{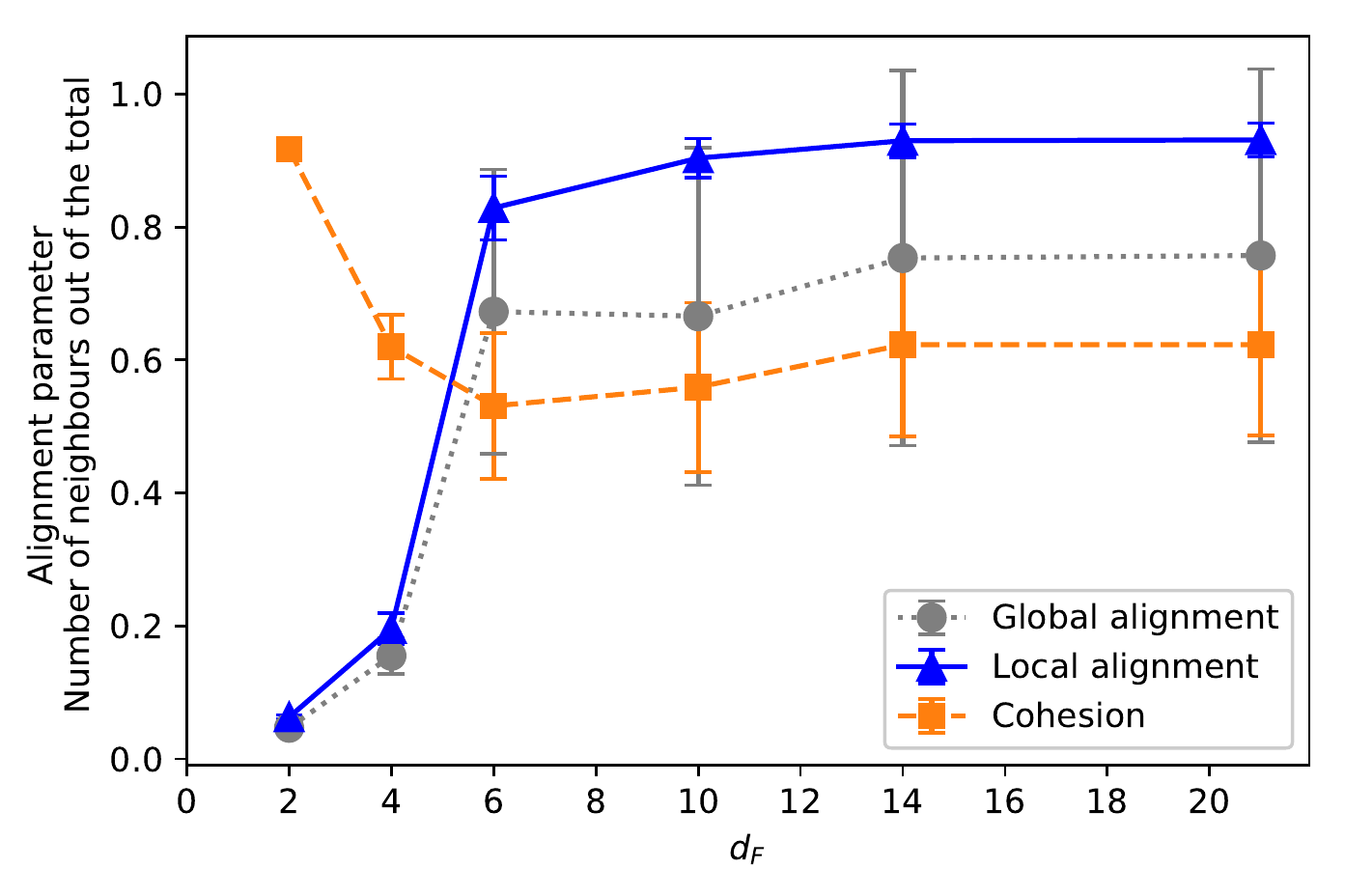}
\caption{Average number of neighbors (in percentage), global and local alignment parameter as a function of the distance $d_F$ to the point where food is placed during the training. Each point is the average of the corresponding parameter over all interaction rounds ($50$) of one trial, and over $100$ trials. $20$ already trained ensembles are considered.}\label{comparison all df}
\end{figure}

Finally, we compare how the alignment and cohesion of the swarms change as a function of the distance at which the resource is placed in the training. Figure~\ref{comparison all df} shows the average local and global alignment parameters, together with the average number of neighbors (at the end of the training) as a function of the distance $d_F$ with which the ensembles were trained. We observe that the farther away the resource is placed, the more strongly the agents align with their neighbors (local alignment) in order to reach it. This is directly related to the individual responses analyzed in Fig.~\ref{comparison_matrixdf}, where one can see that for $d_F\geq 6$ the agents react to positive and negative flow by aligning themselves with their neighbors. Specifically, the observed collective dynamics can be explained in terms of individual responses as follows. The probability of turning around when there is a negative flow and there are not a lot of neighbors (orange-diamonds curve in Fig.~\ref{comparison_matrixdf}) becomes higher as the $d_F$ increases, from $\simeq 0.3$ at $d_F=6$ to $0.6$ at $d_F=21$. The change in the other individual alignment responses (in particular, the other curves in Fig.~\ref{comparison_matrixdf}) is not so large in the region where $d_F>6$, which suggests that the increase in the local alignment and cohesion we observe for $d_F>6$ is mostly due to the strength of the tendency the agents have to turn around when there is a negative flow, even when there are not a lot of neighbors. In addition, the lower values of the global alignment parameter observed in the grey (circles) curve in Fig.~\ref{comparison all df} for $d_F\geq 6$ correspond to the behavior analyzed in Sec.~\ref{subsec alignment}, where it is shown that strongly aligned swarms split into two groups in some of the trials (see also Fig.~\ref{splitting}). With respect to the average number of neighbors, we observe that almost all the agents are within each other's visual range when $d_F=2$. As $d_F$ increases, swarms become initially less cohesive, but once $d_F>6$, they become strongly aligned and consequently once again more cohesive (see discussion in Sec.~\ref{cohesion section} and also figures~\ref{evol cohesion} and~\ref{app fig local evol} for details).

\subsection{Foraging efficiency}\label{SEC for efficiency}

\begin{figure}[htb!]
\includegraphics[width=3.4in]{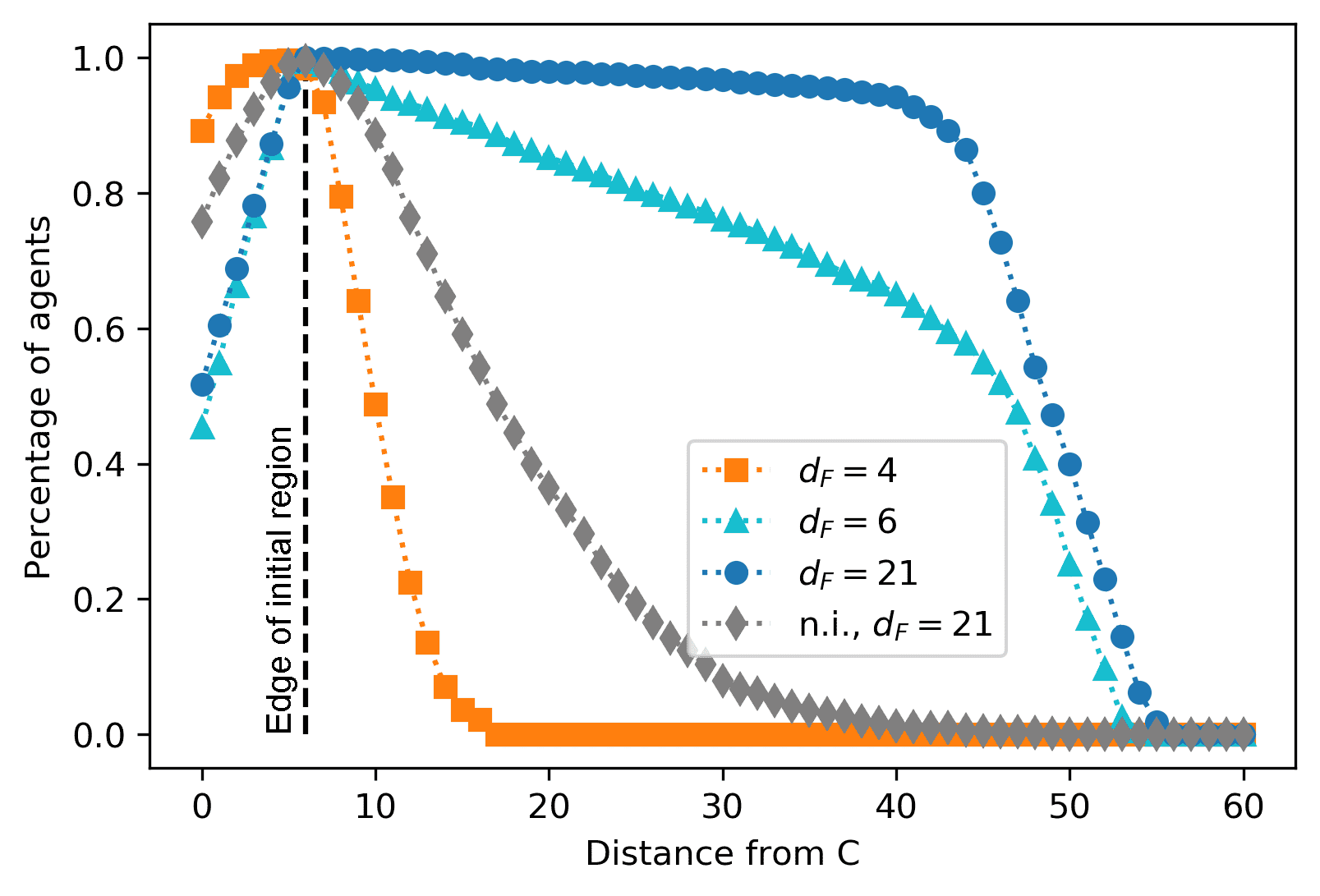}
\caption{Percentage of agents that visit the positions situated at a distance from $C$ given on the horizontal axis (see Fig.~\ref{1d envir}). Since C is located at world position 6, a distance of e.g. 10 on the horizontal axis refers to the world positions 16 and 496. The already trained ensembles walk for one trial of 50 interaction rounds. For each of the four trainings (see legend), the performance of 20 ensembles is considered.}\label{hist efficiency}
\end{figure}

In this section, we study how efficient each type of collective motion is for the purpose of foraging. First, we perform a test where we evaluate how the trained ensembles explore the different world positions. For this test, we analyze which positions in the world are visited by which fraction of agents. The results are given in Fig.~\ref{hist efficiency}. We observe that, for positions within the initial region, agents trained with $d_F=4$ perform better than the others, since they do a random walk that allows them to explore all these positions exhaustively (as evidenced by high percentages of agents that explored positions before the edge of the initial region in Fig.~\ref{hist efficiency}). On the other hand, agents trained with $d_F=6,21$ perform worse when exploring nearby regions, since they form aligned swarms and move straight in one direction. This behavior prevents the agents that are initialized close to the edge of the initial region from exploring the positions inside it. The closer the position is to the edge of the initial region, the more agents visit it because they pass through it when traveling within the swarm. Thus, we conclude that the motion of these swarms is not the optimal to exploit a small region of resources that are located close to each other (a patch). 

Non-interacting (n.i.) agents trained with $d_F=21$ perform slightly better at the intermediate distances than agents trained with $d_F=4$, since they typically travel five steps in a straight line before being randomly reoriented, thereby covering an expected total of 16 positions in one trial (see Sec.~\ref{learning diff scen}). Both curves (grey diamonds and orange squares) show a faster decay in this region than the other two cases ($d_F=6,21$), which is due to the fact that agents do not walk straight for long distances in these two types of dynamics, since they do not stabilize themselves by aligning.

Agents trained with $d_F=21$ reach the best performance for longer distances. In particular, their performance is always better than the performance of agents trained with $d_F=6$, showing that the strategy developed by agents trained with $d_F=21$, namely strong alignment, is the most efficient one for traveling long distances (distance from patch to patch). Agents trained with $d_F=6$ do not align as strongly (see local alignment curve in Fig.~\ref{comparison all df}) and there are more agents that leave the swarm before reaching the furthest positions (see also Fig.~\ref{splitting}), which explains the lower performance at intermediate/long distances (the light blue curve (triangles) has a linear decrease that is stronger in this region than the dark blue curve (circles)). Note that the maximum distance reached by agents is 56; this is simply due to the fact that each trial lasts 50 rounds and the initial positions are within $C\pm6$ (see Fig.~\ref{1d envir}).

In addition, we study the swarm velocity for the different types of collective motions. To do so, we compute the average net distance traveled per round. Considering that the swarm walks for a fixed number of rounds\footnote{The maximum distance agents can travel is 50 because they move at a fixed speed of 1 position per round.} ($50$), we define the normalized swarm velocity as,
\begin{equation}
\langle \xi \rangle= \frac{1}{N}\sum^N_{i=1} \frac{s_i}{50},
\end{equation}
where $N$ is the number of agents and $s_i$ is the net distance traveled by the $i$th agent from the initial position ($x_{i,\,(r=1)}$) to the final position after $50$ interaction rounds ($x_{i,\,(r=50)}$), that is,
\begin{align}
s_i=\min (&(x_{i,\,(r=50)}-x_{i,\,(r=1)})\text{mod} \,W, \nonumber \\
&(x_{i,\,(r=1)}-x_{i,\,(r=50)})\text{mod} \,W),
\end{align}
where $r$ stands for interaction round and $W$ is the world size.

\begin{figure}[h!]
\includegraphics[width=3 in]{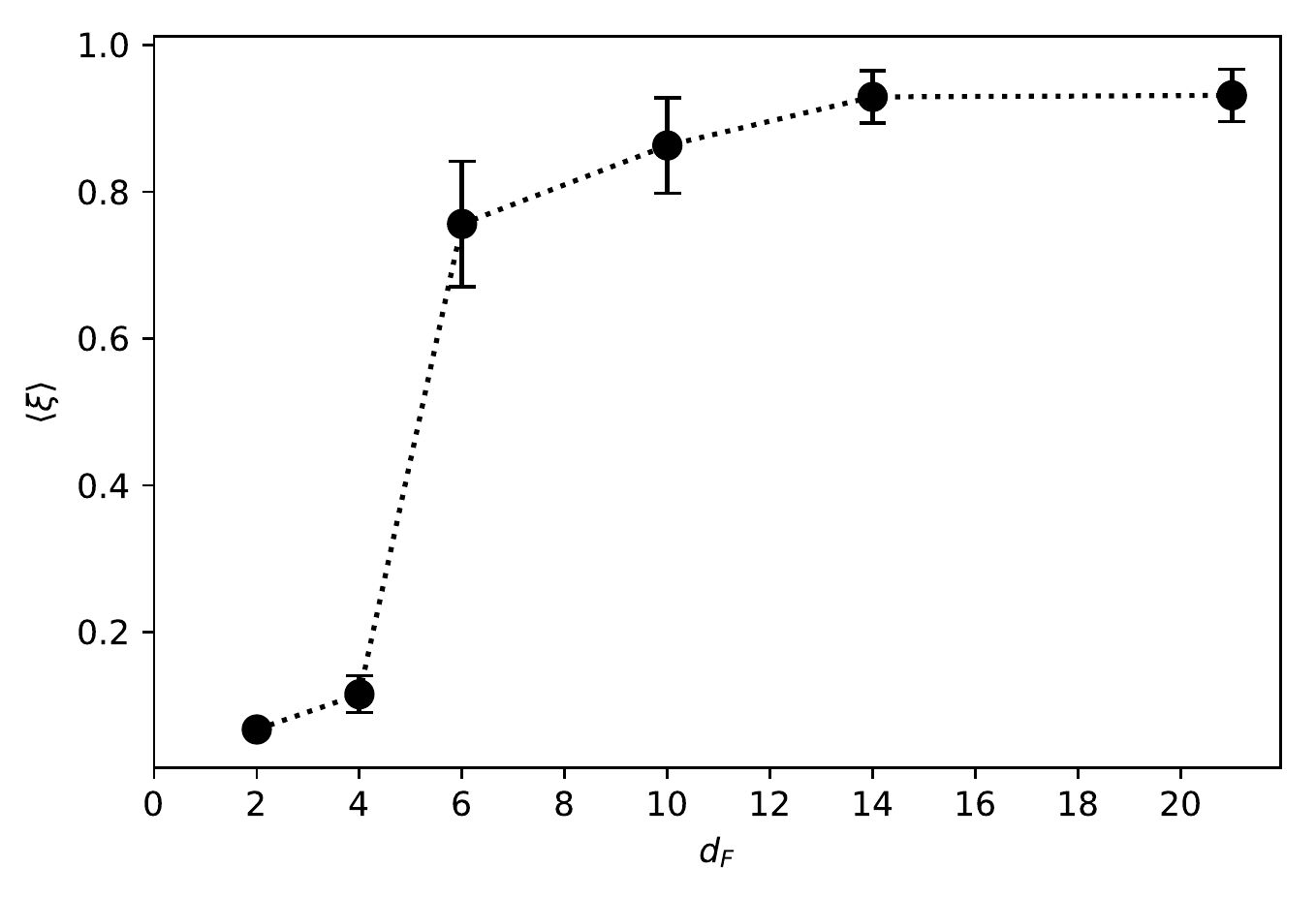}
\caption{Swarm velocity $\langle \xi \rangle$ as a function of the training distance $d_F$. Each point is the average over the agents of 20 independently trained ensembles that have performed 50 independent trials each.}\label{efficiency all df}
\end{figure}

Figure~\ref{efficiency all df} displays the swarm velocity as a function of the distance $d_F$ at which food was placed during the learning process. Agents trained to find distant resources (e.g. $d_F=14,21$) are able to cover a distance almost as large as the number of rounds for which they move. However, while the ensembles trained to find nearby resources (e.g. $d_F=2,4$) form very cohesive swarms, they are less efficient in terms of net distance traveled per interaction round. We observe that the transition between the two regimes happens at $d_F=6$ ---corresponding to the end of the initialization region---, which is consistent with the transitions observed in figures~\ref{comparison_matrixdf} and~\ref{comparison all df} (see discussion in appendix~\ref{APP transition} for more details).

\section{Analysis of the trajectories}\label{SEC Analysis of the learned dynamics}

In this section, we analyze the individual trajectories that result from the different types of swarm dynamics. In order to gather enough statistics, we consider ensembles of agents that have been trained under various conditions, as described above, and let them walk for longer trials so that the individual trajectories are long enough to obtain reliable results. During this process, the agents do not learn anything new anymore; that is, the agents' ECMs remain as they are at the end of the training. Thus, we study the trajectories that emerge from the behavior at the end of the learning process, which can be interpreted as the behavior developed on the level of a population in order to adapt to given evolutionary pressures. The individuals' capacity for learning does not play a role in this analysis.

We focus on the two most representative types of swarms we have observed, i.e. the swarms that emerge from the training with close resources (e.g. $d_F=4$), characterized by strong cohesion; and the swarms that result from the training with distant resources (e.g. $d_F=21$), characterized by strong alignment. For easier readability, in the following we will refer to the swarms formed by agents trained with $d_F=4$ as \textit{cohesive swarms}, and to the swarms formed by agents trained with $d_F=21$ as \textit{aligned swarms}. 

\begin{figure}[h]
\includegraphics[width=3.4in]{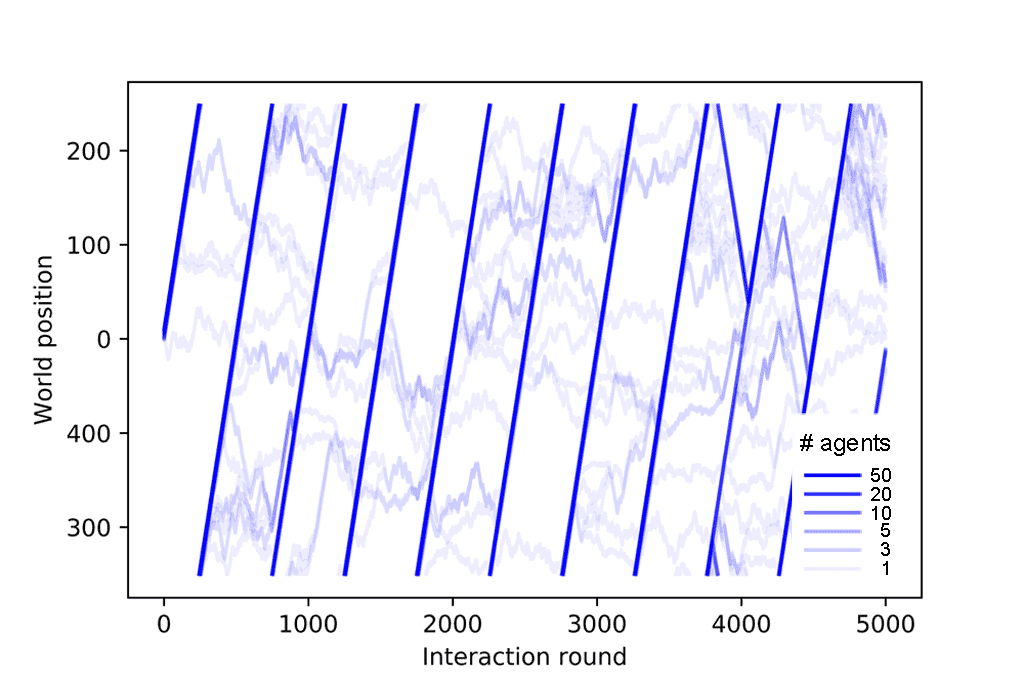}
\caption{Trajectories of one ensemble of 60 agents that were trained with $d_{F}=21$. The world size is $W=500$. Color intensity indicates the number of agents following the same trajectory, i.e. moving within the swarm. Some agents leave the swarm and then rejoin it when the swarm completes the cycle and starts a new turn. Only the first $5000$ interaction rounds (of a total of $10^5$) are shown.}\label{traj levy}
\end{figure}

In the simulations for this analysis, we let each ensemble of agents perform $10^5$ interaction rounds\footnote{Note that each agent moves one position per interaction round.} in a world of size $W=500$ and analyze the individual trajectories. An example of such individual trajectories for the case of agents trained with $d_F=21$ is given in figure~\ref{traj levy}. We observe that some agents leave the swarm at certain points; however, due to the 'closed' nature of our world model, they have the possibility of rejoining the swarm once it completes a cycle and starts a new turn around the world. Due to these environmental circumstances, the agents exhibit two movement modes: when they are alone and when they are inside the swarm. By looking at Fig.~\ref{traj levy}, one can see how agents exhibit directional persistence when they move within the swarm, since they have learnt to align themselves with their neighbours as a strategy for stabilizing their orientations. However, trajectories become more tortuous as agents leave the swarm and walk on their own. Note that it is only possible for individuals to leave the swarm\footnote{The specific probabilities of doing so are given in Fig.\ref{Learned-probabilities} (a) and analyzed in Sec.\ref{SEC ind responses}.} because of the weaker cohesion exhibited by aligned swarms (see Sec.~\ref{cohesion section}). This bimodal behavior can occur in nature (see e.g. collective motion and phase polyphenism in locusts \citep{Ariel15,Pener09}), where individuals may benefit from collective alignment, for instance, to travel long distances in an efficient way, but they move independently to better explore nearby resources (see Sec.\ref{SEC for efficiency} for details on exploration efficiency of the different collective dynamics).

In the following sections, we characterize the trajectories and assess how well the agents' movement patterns fit to well-known foraging models such as L\'evy walks or composite correlated random walks. 

\subsection{Theoretical foraging models}\label{for models}
This work is directly related to foraging theory, since the task we set for the learning process is to find food in different environmental conditions. For this reason, we will analyze our data to determine whether the movement patterns that emerge from this learning process support any of the most prominent search models. For environments with scarce resources (e.g. patchy landscapes), these models are the L\'evy walks \citep{Viswanathan99} and the composite correlated random walks (CCRW) \citep{Benhamou92}. 

In order to analyze the trajectories and determine which type of walk fits them best, the distribution of step lengths is studied, where a \textit{step length} is defined as the distance between two consecutive locations of an organism. Intuitively, the optimal strategy for navigating a patchy landscape allows for both an exhaustive exploration inside patches and an efficient displacement between patches, employing some combination of short and long steps. L\'evy walks have a distribution of step lengths in which short steps have higher probability of occurrence but arbitrarily long steps can also occur due to its power-law (PL) tail. In two- and three-dimensional scenarios, the direction of motion is taken from a uniform distribution from 0 to $2\pi$, which implies that L\'evy walks do not consider directionality in the sense of correlation in direction between consecutive steps \citep{Pyke15}. On the other hand, CCRW and the simpler version thereof, composite random walks (CRW), consist of two modes, one intensive and one extensive, which are mathematically described by two different exponential distributions of the step lengths. The intensive mode is characterized by short steps (with large turning angles in 2D) to exploit the patch, whereas the extensive mode ---whose distribution has a lower decay rate--- is responsible for the inter-patch, straight, fast displacement. CCRW in addition allow for correlations between the directions of successive steps.

Even though the models are conceptually different, the resulting trajectories may be difficult to distinguish \citep{Benhamou07,Plank08,Plank09}, even more if the data is incomplete or comes from experiments where animals are difficult to track. In the past years, many works have been published that try to provide techniques to uniquely identify L\'evy walks \citep{Reynolds12,Humphries13,Gautestad12} and to differentiate between the two main models \citep{Benhamou07,Jansen12,AuguerMethe15}. For instance, some of the experiments that initially supported the hypothesis that animals perform L\'evy walks \citep{Viswanathan96,Sims08,deJager11} were later reanalyzed to support the conclusion that more sophisticated statistical techniques are, in general, needed \citep{Edwards07,Edwards11,Edwards12,Jansen12}. Apart from that, there exist several studies that relate different models of collective dynamics to the formation of L\'evy walk patterns under certain conditions \citep{Santos09,Reynolds16}. For instance, it has been shown \citep{Reynolds13swarmlevy} that L\'evy walk movement patterns can arise as a consequence of the interaction between effective leaders and a small group of followers, where none of them has information about the resource.

\begin{figure}[h]
\includegraphics[width=3.4in]{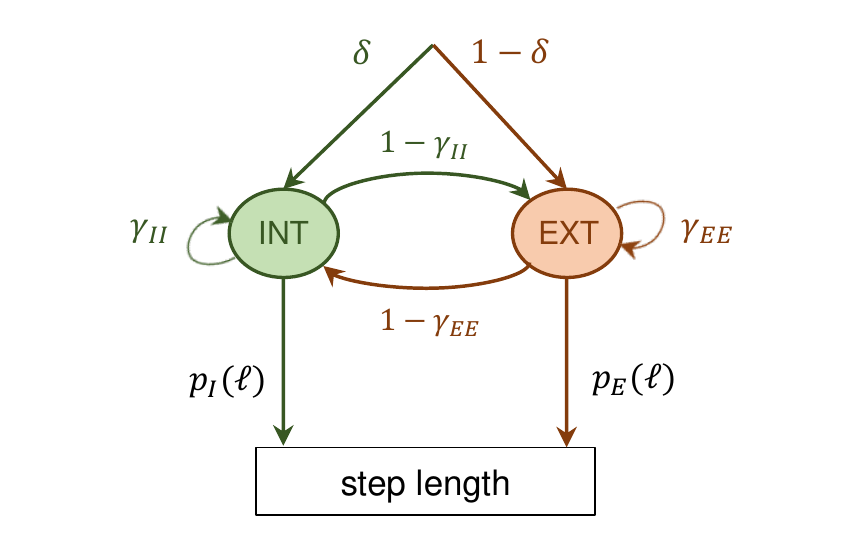}
\caption{Hidden Markov model for the CCRW. There are two modes, the intensive and the extensive, with probability distributions given by $p_I$ and $p_E$ (see text for details). The probability of transition from the intensive (extensive) to the extensive (intensive) mode is given by $1-\gamma_{II}$ ($1-\gamma_{EE}$), where $\gamma_{II}$ and $\gamma_{EE}$ are the probabilities of remaining in the intensive and extensive mode respectively. $\delta$ is the probability of starting in the intensive mode.}\label{hmm}
\end{figure}

In our study, we consider the three models we have already mentioned (PL, CCRW and CRW), together with Brownian motion (BW) as a baseline for comparison. Since our model is one-dimensional, a distribution of the step lengths is sufficient to model the trajectories we observe, and no additional distributions, such as the turning angle distributions, are needed. In addition, the steps are unambiguously identified: a step has length $\ell$ if the agent has moved in the same direction for $\ell$ consecutive interaction rounds. Finally, since space in our model is discretized, we consider the discrete version of each model's probability density function (PDF). More specifically, the PDFs we consider are,

\begin{enumerate}

\item Brownian motion (BW):
\begin{equation}
p\,(\ell)=(1-e^{-\lambda})\,e^{-\lambda (\ell-1)},\,\,\,\, \ell\geq1,\label{eq BW}
\end{equation} 
where $\lambda$ is the decay rate and the minimum value a step length can have is, in our case, known to be $1$, since agents move at a constant speed of one position per interaction round.

\item Composite random walk (CRW):
\begin{align}
p\,(\ell)&=p\,(1-e^{-\beta_I})\, e^{-\beta_I(\ell-1)}\nonumber\\&  
+(1-p)\,(1-e^{-\beta_E})\,e^{-\beta_E(\ell-1)}, \,\,\,\, \ell\geq1,   \label{eq CRW} 
\end{align}
where $p$ is the probability of taking the intensive mode, $\beta_I$ is its decay rate and $\beta_E$ is the decay rate of the extensive mode. In this case, again, the minimum step length is $1$.

\item Composite correlated random walk (CCRW):
\begin{align}
p_I\,(\ell|I)&=(1-e^{-\lambda_I})\,e^{-\lambda_I (\ell-1)},\,\,\,\, \ell\geq1,\label{eq CCRW pi}\\ 
p_E\,(\ell|E)&=(1-e^{-\lambda_E})\,e^{-\lambda_E (\ell-1)},\,\,\,\, \ell\geq1,\label{eq CCRW pe} \\
p(m'=E|m=I)&=1-\gamma_{II},\\
p(m'=I|m=E)&=1-\gamma_{EE},
\end{align}
where $p_I\,(\ell|I)$ and $p_E\,(\ell|E)$ are the PDFs of the step lengths $\ell$ corresponding to the intensive and extensive mode respectively. Denoting the mode in which the agent is as $m$ and the mode to which the agent transitions as $m'$, $p(m'=E|m=I)$ is the transition probability from the intensive to the extensive mode and $p(m'=I|m=E)$, from the extensive to the intensive mode. $\lambda_I, \lambda_E, \gamma_{II}$ and $\gamma_{EE}$ are parameters of the model. The main difference between the CRW and the CCRW models is that, in the latter, the step lengths are correlated, i.e. the order of the sequence of step lengths, and thus the order in which the movement modes alternate, matters. The CCRW is modeled as a hidden Markov model (HMM) (see \citep{AuguerMethe15,Zucchini09}) with two modes, the intensive and the extensive. Figure~\ref{hmm} shows the details of the model and the notation for the transition probabilities between modes.

\item Power-law (PL):
\begin{equation}
p\,(\ell)=\frac{\ell^{-\mu}}{\zeta(\mu,1)},\,\,\,\, \ell\geq 1,\label{eq PL}
\end{equation}
where the normalization factor $\zeta(\mu,1)=\sum_{a=0}^{\infty} (a+1)^{-\mu}$ is the Hurwitz zeta function \citep{Clauset09}. The parameter $\mu$ gives rise to different regimes of motion: L\'evy walks are characterized by a heavy-tailed distribution, with exponents $1< \mu \leq 3$, which produces superdiffusive trajectories, whereas $\mu>3$ corresponds to normal diffusion, as exhibited by Brownian walks. We note that the above distribution starts at $\ell=1$, which is the shortest possible distance that our agents move in a straight line. The scale of this minimum step length is determined by the embodied structure of the organism and is typically considered to be one body length \citep{Pyke15}. Some other works (e.g. \citep{Clauset09,Humphries13}) consider a variant of the above distribution that only follows the PL form for steps longer than some threshold $\ell_0$, for example when analysing experimental data that become increasingly noisy at short step-lengths. However, since the step lengths resulting from our simulations are natively discrete, the unbounded PL distribution given in eq.~\eqref{eq PL} seems appropriate. Moreover, if one were to introduce a lower bound $\ell_0>1$, one would need to add more parameters in the model to account for the probabilities $p\,(\ell)$ for all $1\le\ell<\ell_0$, which we consider an unnecessary complication. This is particularly relevant when it comes to comparing PL to BW, CRW or CCRW as models for fitting our data: since none of the other models include lower bounds, we achieve a more consistent comparison by a parsimonious approach that includes all step lengths $\ell\geq 1$ in the PL model and thereby abstains from additional free parameters. 
\end{enumerate}

\subsection{Visual analysis}
In this section, we study the general characteristics of the trajectories of both types of swarm dynamics. We start by analyzing how diffusive the individual trajectories are depending on whether the agents belong to an ensemble trained with $d_F=21$ (dynamics of aligned swarms) or $d_F=4$ (dynamics of cohesive swarms). More specifically, we analyze the mean squared displacement (MSD), defined as,
\begin{equation}
\langle \delta r ^2 \rangle =\langle |x(t)-x_0|^2 \rangle,
\end{equation}
where $x_0$ is the reference (initial) position and $x(t)$ is the position after time $t$ elapsed. In general, the MSD increases with the time elapsed as $\langle \delta r ^2 \rangle \sim t^\alpha$. Depending on the exponent $\alpha$, the diffusion is classified as normal diffusion ($\alpha=1$), subdiffusion ($\alpha<1$) or superdiffusion ($\alpha >1$), which is called ballistic diffusion when $\alpha=2$. For instance, a Brownian particle undergoes normal diffusion, since its MSD grows linearly with time. 

\begin{figure}[htb!]
\includegraphics[width=3.4in]{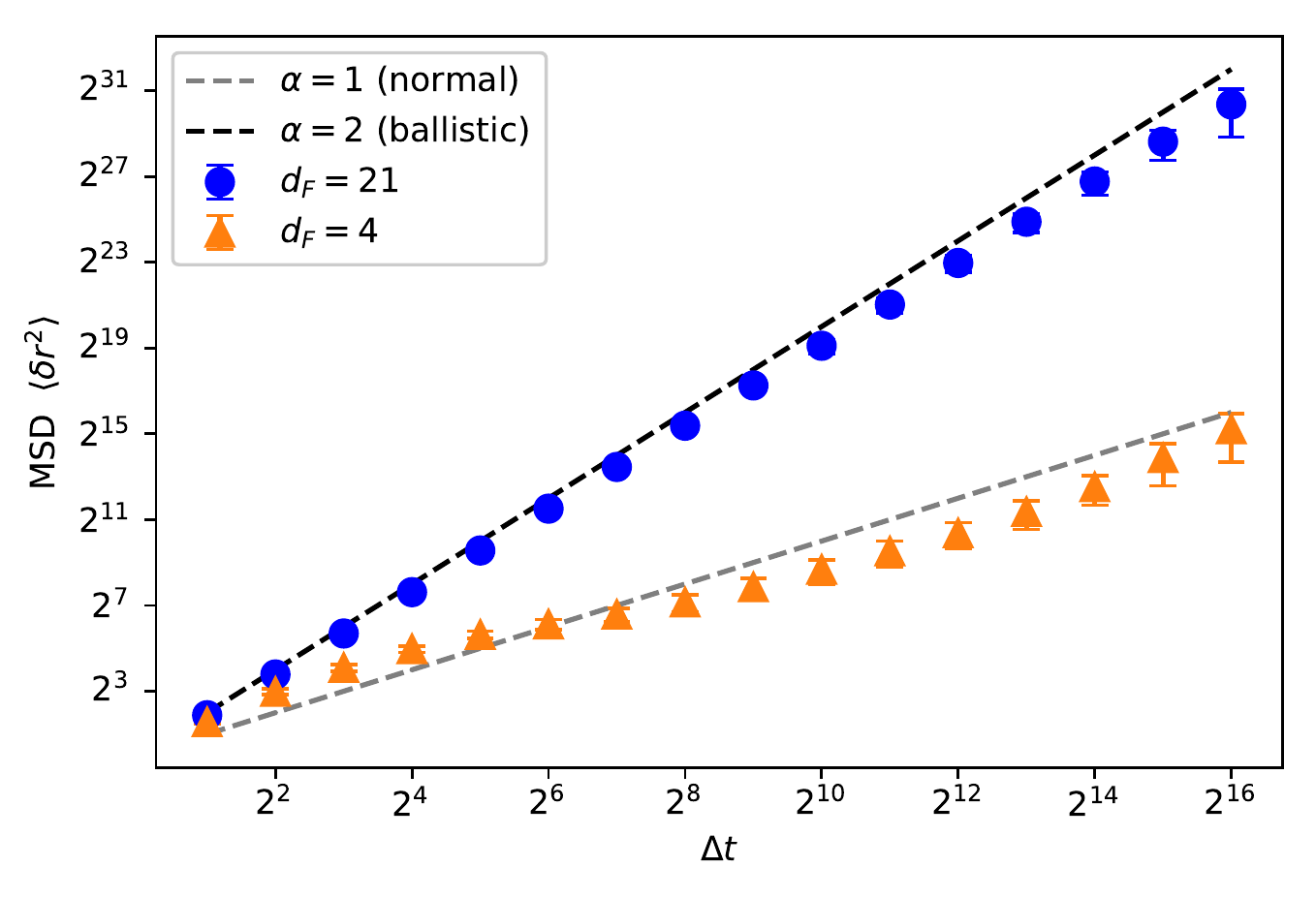}
\caption{Log-log (base 2) plot of the MSD as a function of the time interval for two types of trajectories: trajectories performed by agents trained with $d_F=21$ (blue curve, circles) and by agents trained with $d_F=4$ (orange curve, triangles).  We observe that the former present ballistic diffusion, whereas the latter exhibit close-to-normal diffusion. 600 individual trajectories (10 ensembles of 60 agents) are considered for each case.}\label{msd}
\end{figure}

Figure~\ref{msd} shows that the dynamics of aligned swarms leads to superdiffusive individual trajectories (ballistic, with $\alpha=2$), whereas the trajectories of agents that belong to cohesive swarms exhibit close-to-normal diffusion. The anomalous diffusion (superdiffusion) exhibited by the agents trained with $d_F=21$ (curve with blue circles in Fig.~\ref{msd}) favors the hypothesis that the swarm behavior may induce L\'evy-like movement patterns, since L\'evy walks are one of the most prominent models describing superdiffusive processes. However, CCRW can also produce superdiffusive trajectories \citep{Benhamou92,Benhamou07}. In contrast, agents trained with $d_F=4$ do not align with each other and the normal diffusion shown in Fig.~\ref{msd} is indicative of Brownian motion. 

\begin{figure}[htb!]
\subfigure[Trajectory of an agent that belongs to an aligned swarm.]{\includegraphics[width=3.4in]{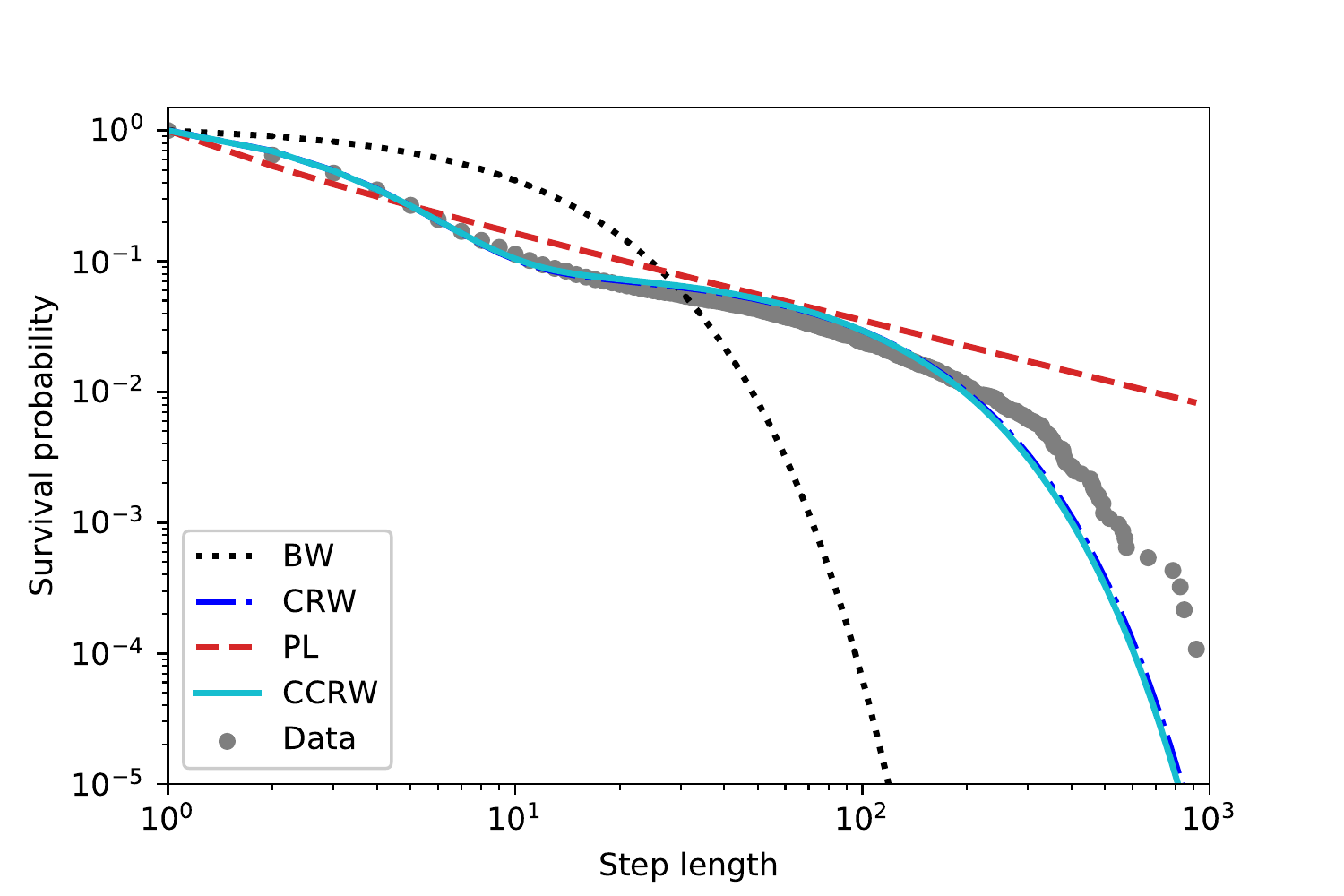}}
\subfigure[Trajectory of an agent that belongs to a cohesive swarm.]{\includegraphics[width=3.4in]{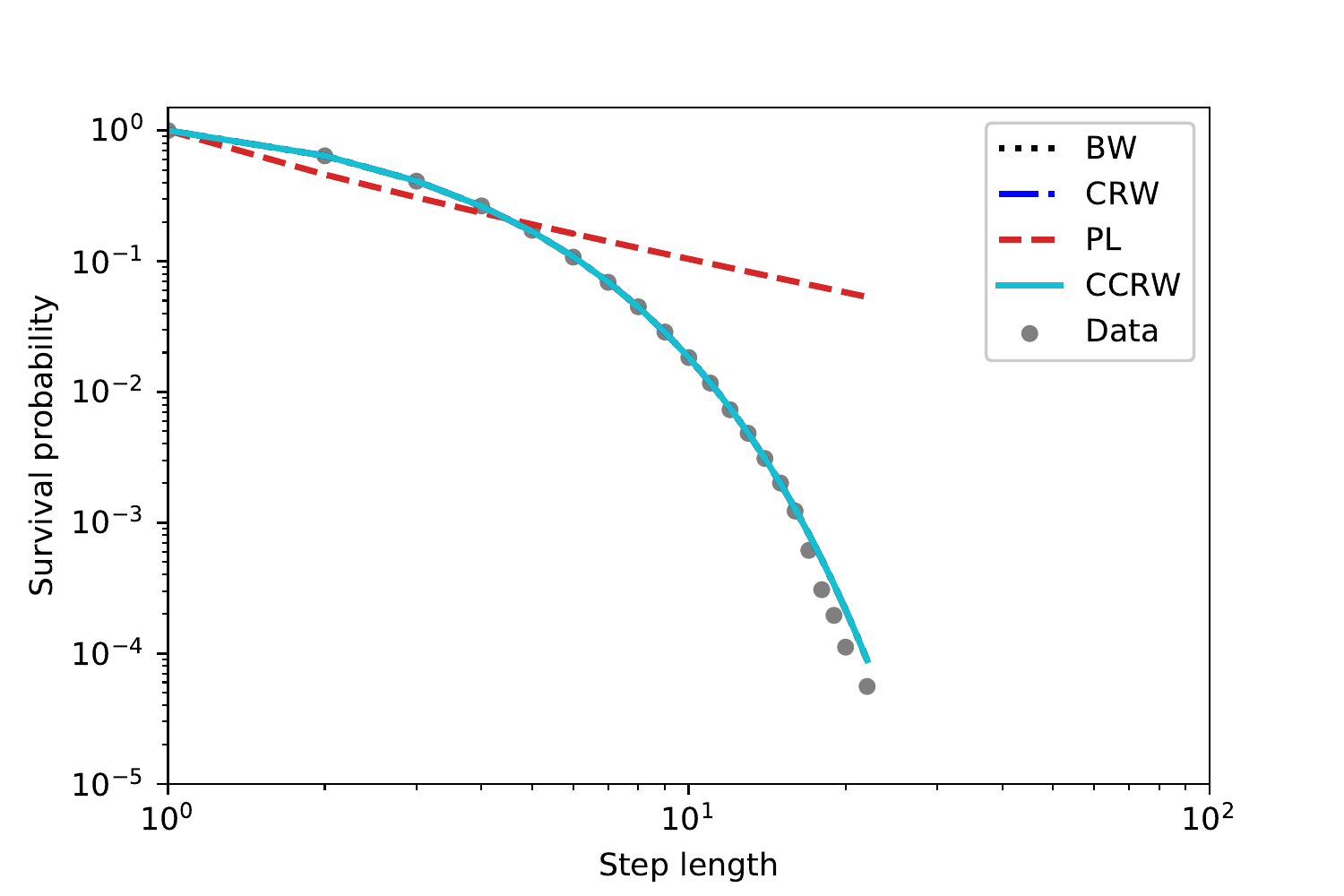}}
\caption{Survival probability (percentage of step lengths larger than the corresponding value on the horizontal axis) as a function of the step length. Each panel depicts the data from the trajectory of one agent picked from (a) aligned swarms and (b) cohesive swarms, so that this figure represents the most frequently observed trajectory for each type of dynamics. The survival distributions of the four candidate models are also plotted. The distributions for each model are obtained considering the maximum likelihood estimation of the corresponding parameters (see Sec.~\ref{stat results} for details). The curve for the CCRW model is obtained by an analytic approximation of the probabilities of each step length, given the maximum likelihood estimation of its parameters. Since the order of the sequence of step lengths is not relevant for this plot, we estimate the probabilities of each step length as $p\left(\ell\right)=p'\left(1-e^{-\hat{\lambda}_{I}}\right)e^{-\hat{\lambda}_{I}\left(\ell-1\right)}+\left(1-p'\right)\left(1-e^{-\hat{\lambda}_{E}}\right)e^{-\hat{\lambda}_{E}\left(\ell-1\right)}$ (see eq.~\eqref{eq CRW}) with $p' \simeq \frac{1}{1-\gamma_{II}}\left(\frac{1}{1-\gamma_{II}}+\frac{1}{1-\gamma_{EE}}\right)^{-1}$.  }\label{survival}
\end{figure}

The analysis presented above already shows a major difference between the two types of swarm dynamics but it is in general not sufficient to determine which theoretical model (L\'evy walks or CCRW) best fits the data from aligned swarms. According to \citep{Benhamou07}, one possible way to distinguish between composite random walks and L\'evy walks is to look at their survival distributions, which is the complement of the cumulative distribution function, giving the fraction of steps longer than a given threshold. L\'evy walks would exhibit a linear log-log relationship when this type of distribution is plotted, whereas CCRW exhibit a non-linear relation. Figure~\ref{survival} compares the survival distributions of two trajectories, one from each type of swarm, to those predicted by the best-fitting models of each of the four classes. The maximum length observed in the $d_F=4$-trajectory is of the order of $10$, whereas in the case of the $d_F=21$-trajectory, it is one order of magnitude larger. The most prominent features one infers from these figures are that all models except PL seem to fit the data of the $d_F=4$-trajectory, and that Brownian motion is clearly not a good model to describe the $d_F=21$-trajectory. In addition, Fig.~\ref{survival} (a) is curved and seems to be better fit by the CCRW. However, when other trajectories of agents trained with $d_F=21$ are plotted in the same way, we see that data seems to better follow the straight line of the PL rather than the CCRW (see for example Fig.~\ref{app survival plot}).

While visual inspection may be an intuitive way of assessing model fit, and one that is easy to apply at small scales, it would be preferable to use a method that yields quantitative and objective, repeatable assessments of how well various models fit a given data set. Moreover, we generated 600 individual trajectories per type of swarm, in order to support statistically meaningful conclusions, and at this scale visual inspection quickly becomes infeasible. For this reason, we now turn to a more rigorous statistical analysis of the individual trajectories.

\subsection{Statistical analysis}\label{stat results}
In order to determine which of the mentioned models best fits our data, we perform the following three-step statistical analysis for each individual trajectory: (i) first, we optimize each family of models to get the PDF that most likely fits our data via a maximum likelihood estimation (MLE) of the model parameters. (ii) Then, we compare the four different candidate models among them by means of the Akaike information criterion (AIC) \citep{Burnham04} and (iii) finally, we apply an absolute fit test for the best model. We repeat this analysis for agents trained with $d_F=4$ and $d_F=21$, yielding a total of 600 individual trajectories per type of training (10 ensembles of aligned swarms and 10 of cohesive swarms, where each ensemble has 60 agents). The simulation of $10^5$ interaction rounds is performed for each ensemble independently. In order to do the statistical analysis, each individual trajectory is divided into steps, which are defined in our case as the distance the agent travels without turning. We obtain sample sizes that range from $4000$ to $17000$ steps for trajectories of agents trained with $d_F=21$ and from $20000$ to $40000$ steps in the case with $d_F=4$.

The following provides more detail on the analysis, starting with the MLE method, which consists in maximizing the likelihood of each model candidate with respect to its parameters. The likelihood function is generally defined as,
\begin{equation}
\mathcal{L}(\theta| \ell_{i=1..S})=\prod_{i=1}^{S} p\,(\ell_i,\theta),\label{likelihood function}
\end{equation}
where $S$ is the sample size and $p(\ell_i,\theta)$ is the PDF of the given model ---that depends on the model parameters $\theta$--- evaluated at the data point $\ell_i$. Details on the maximization process and the computation of the likelihood function in the case of CCRW, which is more complicated since consecutive step lengths $\ell_i$ are not sampled independently, are given in appendix~\ref{appendix mle}. In the following, we denote the values of the parameters that maximize the likelihood and the value of the maximum likelihood with hatted symbols.

Table~\ref{tabla:mle par} shows the MLE parameters we have obtained for each model and for each swarm type. We observe that, in the $d_F=21$ case, the decay rates of the exponential distributions ($\hat{\lambda},\hat{\beta}_E,\hat{\lambda}_E$) are very small (approx. of the order of $0.01$) compared to the decay rates in the $d_F=4$ case (approx. of the order of $0.3$), which implies that the former allows for longer steps to occur with higher probability. The decay rates of the intensive modes ($\hat{\beta}_I,\hat{\lambda}_I$) are comparable to the BW decay rate of $d_F=4$ because they account for the shorter, more frequent steps, which occur in both types of dynamics ---in the $d_F=21$ case, agents perform shorter steps when they leave the swarm and move on their own---. Also note that the power-law coefficient $\mu\simeq 1.6$ in the $d_F=21$ case implies that the PL model is that of a L\'evy walk.

\begin{table}[htb]
\begin{tabular}{|l|c|c|c|}
\hline
\multicolumn{2}{|c|}{Model} & $d_F=21$ & $d_F=4$ \\ \hline \hline
BW & $\hat{\lambda}$ & 0.083 $\pm$ 0.032 & 0.305 $\pm$ 0.052 \\ \hline \hline
\multirow{3}{1cm}{CRW} & $\hat{\beta}_I$ & 0.37 $\pm$ 0.15 & 0.6 $\pm$ 1.1 \\ \cline{2-4} 
& $\hat{\beta}_E$ & 0.0126 $\pm$ 0.0069 & 0.302 $\pm$ 0.048 \\ \cline{2-4} 
& $\hat{p}$ &  0.879 $\pm$ 0.040 & 0.196 $\pm$ 0.059 \\ \hline \hline  
PL & $\hat{\mu}$ &  1.59 $\pm$ 0.10 & 1.657 $\pm$ 0.066 \\ \hline \hline
\multirow{5}{1cm}{CCRW} & $\hat{\delta}$ & 0.23 $\pm$ 0.31 & 0.12 $\pm$ 0.13 \\ \cline{2-4}
& $\hat{\lambda}_I$ & 0.37 $\pm$ 0.15 & 0.36 $\pm$ 0.21 \\ \cline{2-4}
& $\hat{\lambda}_E$ & 0.0134 $\pm$ 0.0066 & 0.300 $\pm$ 0.047 \\ \cline{2-4}
& $\hat{\gamma}_{II}$ & 0.839 $\pm$ 0.078 & 0.749 $\pm$ 0.060 \\ \cline{2-4}
& $\hat{\gamma}_{EE}$ & 0.013 $\pm$ 0.020 & 0.09 $\pm$ 0.26 \\ \hline
\end{tabular}
\caption{Average values of the MLE parameters for the different models. 600 trajectories are analyzed for each type of swarm.}
\label{tabla:mle par}
\end{table}

Once the value of the maximum likelihood $\mathcal{\hat{L}}$ is obtained for each model, it is straightforward to compute its Akaike value,
\begin{equation}
AIC=2k-2\ln(\mathcal{\hat{L}}),
\end{equation}
where $k$ is the number of parameters of the model. The model with the lowest $AIC$ ($AIC_{min}$) is the best model (out of the ones that are compared) to fit the data \citep{Burnham04}. In order to compare the models in a normalized way, the Akaike weights are obtained from the Akaike values as,
\begin{equation}
w_i=\frac{e^{-\frac{1}{2}\Delta_i (AIC)}}{\sum_{k=1}^K e^{-\frac{1}{2}\Delta_k (AIC)}}
\end{equation}
where $w_i$ is the Akaike weight of the $i$th model, and $\Delta_i(AIC)=AIC_i-AIC_{min}$, with $AIC_i$ the Akaike value of the $i$th model. The interpretation of $w_i$ is not straightforward but, as it was argued in \citep{Symonds11}, "Akaike weights can be considered as analogous to the probability that a given model is the
best approximating model". $K$ is the total number of models under comparison, so that the Akaike weights are normalized as $\sum_{i=1}^K w_i=1$. In appendix~\ref{APP tables}, we present detailed tables with the results of this statistical analysis for three trajectories, two for training with $d_F=21$ and one with $d_F=4$.

\begin{figure}[htb!]
\subfigure[Akaike weights of the 600 trajectories of agents trained with $d_F=21$.]{\includegraphics[width=3.4in]{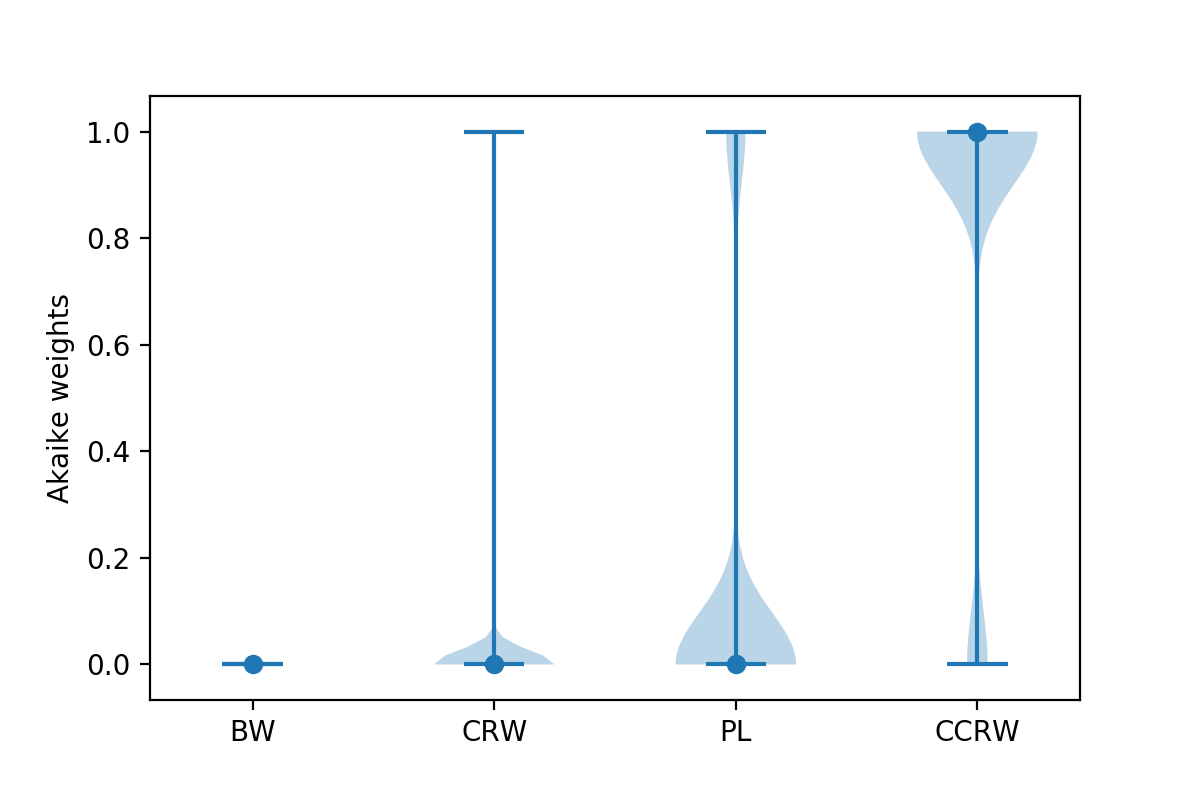}}
\subfigure[Akaike weights of the 600 trajectories of agents trained with $d_F=4$.]{\includegraphics[width=3.4in]{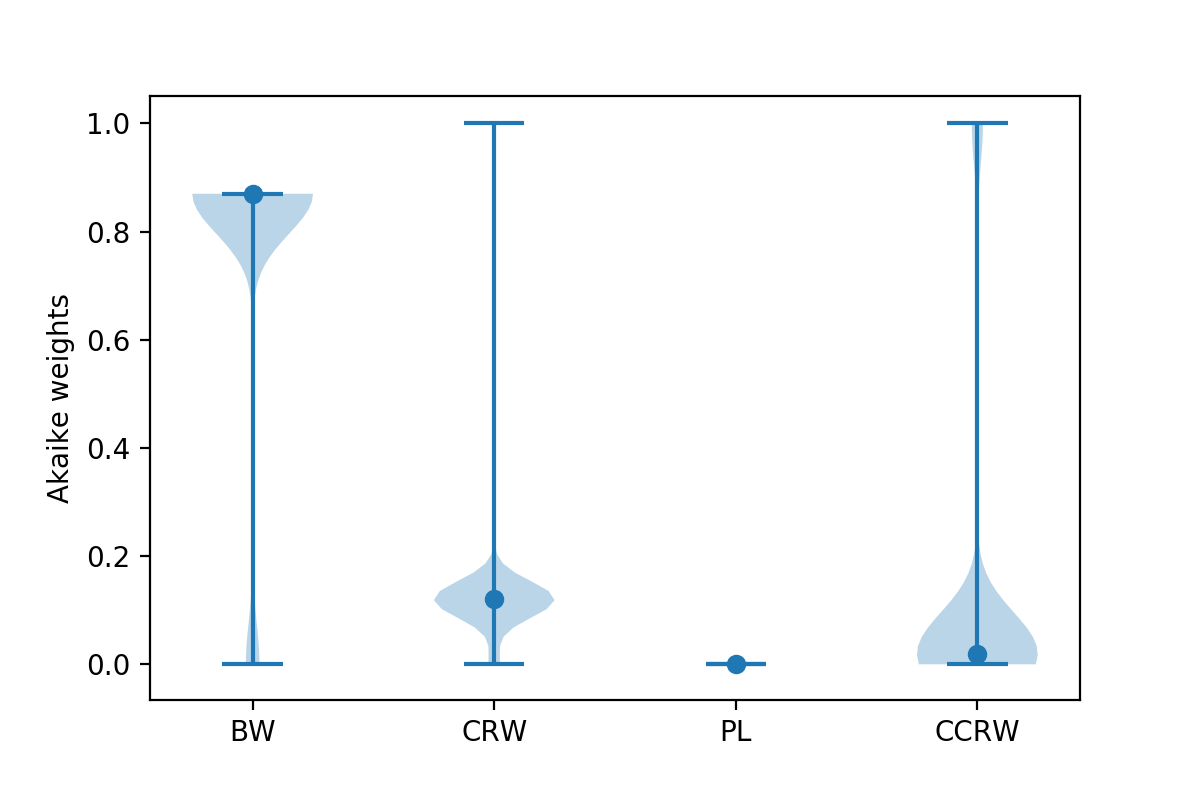}}
\caption{Violin plots that represent the Akaike weights obtained for each model for the trajectories of agents trained with (a) $d_F=21$ (aligned swarms) and (b) $d_F=4$ (cohesive swarms). 600 individual trajectories ---per type of swarm--- were analyzed for each plot. The '$\bullet$' symbol represents the median and the vertical lines indicate the range of values in the data sample (e.g. PL model in figure (a) has extreme values of 0 and 1). Shaded regions form a smoothed histogram of the data (e.g. the majority of Akaike weights of the CCRW model in figure (a) have value 1, and there are no values between 0.2 and 0.8). See text for more details.}\label{AICw}
\end{figure}

Figure~\ref{AICw} shows the results of the Akaike weights obtained for each of the 600 trajectories analyzed for each type of swarm. In the case of the aligned swarms (figure~\ref{AICw} (a)), we observe that the BW model is discarded in comparison to the other models, since its Akaike weight is zero for all trajectories. $85\%$ of trajectories have Akaike weight of 1 for the CCRW model\footnote{$0.8\%$ of trajectories have $w_{\text{CRW}}=1$, which is to be expected since the MLE parameters of both CRW and CCRW models are roughly the same.} and 0 for the rest of the models, whereas $14\%$ of trajectories have Akaike weight of 1 for the PL model and 0 for the rest. This result is in agreement with previous works that claim that "selection pressures give CCRW L\'evy walk characteristics" \citep{Reynolds13}. Therefore, the majority of individual trajectories are best fit by CCRW with two exponential distributions whose means are $\hat{\lambda}_I^{-1}\simeq 2.7$ and $\hat{\lambda}_E^{-1}\simeq 75$, which give the movement patterns L\'evy-walk features. In addition, a considerable percentage of trajectories are indeed best fit by a power-law distribution with exponent $\mu=1.6$, that is, a L\'evy walk.

On the other hand, the cohesive swarms (figure~\ref{AICw} (b)) show high Akaike weights for all models except the PL, which implies that only the PL model can be discarded as a description of the observed movement patterns. $92\%$ of trajectories have Akaike weights $0.87$, $0.12$ and $0.01$ for the BW, CRW and CCRW models, respectively. The remaining $8\%$ of trajectories have $w_{\text{CCRW}}=1$. However, note in Table~\ref{tabla:mle par} that the MLE parameters for the four models in fact specify particular limiting cases that correspond to very similar probability distributions, which indicates that the movement has essentially the same characteristics in all models (see also Fig.~\ref{survival} (b)). In particular, the intensive and extensive modes in the CRW and CCRW models are of the same order, which implies that there is effectively one mode. Overall, the type of motion that the agents in these swarms exhibit has Brownian motion characteristics.

Finally, we study the goodness-of-fit (GOF) of the different models. For models that deal with i.i.d. variables (BW, CRW, PL), it is enough to perform a likelihood ratio test, whose $p$-value indicates how well the data is fit by the model. Within our framework, a low $p$-value, namely $p<0.05$, means that the model can be rejected as a description for the observed data with a confidence of $95\%$. The closer $p$ is to 1, the better the model fits the data. In the case of CCRW, a more involved method is needed due to the correlation in the data. Specifically, we first compute the uniform pseudo-residuals (see \citep{Zucchini09}) and then we perform a Kolmogorov-Smirnov (KS) test to check for uniformity of the mid-pseudo-residuals. Details on the methods used in both GOF tests are given in Appendix~\ref{appendix stat}. Even though a visual inspection of figure~\ref{survival} suggests that the CRW, PL, CCRW models fit the data reasonably well, a quantitative analysis gives $p$-values of $p<0.01$ for most of the trajectories fitted by the BW and PL models. Some trajectories fitted by the CRW model give better fittings, e.g. the best $p$-values are $p=0.97$ and $p=0.36$ for a trajectory of an agent trained with $d_F=4$ and $d_F=21$, respectively. In the CCRW case, we give the average value of the KS distance obtained in the 600 trajectories, which is $D_{KS}=0.134\pm 0.016$ and $D_{KS}=0.189\pm 0.046$ for $d_F=4$ and $d_F=21$-trajectories respectively \footnote{Note that a perfect fit gives $D_{KS}=0$ and the worst possible fitting gives $D_{KS}=1$.}. More details on the GOF tests and their results are given in appendix~\ref{gof section}.

A closer inspection reveals that this relatively poor fit is mostly due to irregularities in the tails of the observed distributions. However, more importantly, we note that the trajectories were in fact not drawn from a theoretical distribution chosen for its mathematical simplicity, but result from  individual interactions of agents that have learned certain behaviors. In this regard, with respect to the sometimes low goodness-of-fit values, our simulations lead to similar challenges as the analysis of experimental data from real animals (see e.g. \citep{AuguerMethe15}). Nonetheless, the above analysis does provide a more robust account of key features of the collective dynamics.

\section{Conclusions}\label{SEC summary}
We have studied the collective behavior of artificial learning agents, more precisely PS agents, that arises as they attempt to survive in foraging environments. More specifically, we design different foraging scenarios in one-dimensional worlds in which the resources are either near or far from the region where agents are initialized.

This ansatz differs from existing work in that PS agents allow for a complex, realistic description of the sensory (percepts) and motor (actions) abilities of each individual. In particular, agents can distinguish how other agents within visual range are oriented and if the density of agents is high or low in the front and at the back of their visual area. Based on this information, agents can decide whether to continue moving in their current direction or to turn around and move in the opposite direction. Crucially, there are no fixed interaction rules, which is the main difference that sets our work apart from previous approaches, like the self-propelled particle (SPP) models or other models from statistical physics. Instead, the interactions emerge as a result of the learning process agents perform within a framework of reinforcement learning. The rewards given as part of this learning process play a role analogous to evolutionary pressures in nature, by enhancing the behaviors that led the agent to be rewarded. Therefore, by varying the task and reward scheme and studying the resulting behaviors, our approach allows us to test different \emph{causal explanations} for specific observed behaviors, in the sense of evolutionary pressures proposed to have led to these behaviors.

In this work, we have considered scenarios where the food is situated inside or far from the region where agents are initialized and we have observed that the initially identical agents develop very different individual responses ---leading to different collective dynamics--- depending on the distance they need to cover to reach the reward (food source). Agents learn to form strongly aligned swarms to get to distant food sources, whereas they learn to form cohesive (but weakly aligned) swarms when the distance to the food source is short. 

Since we model each individual as an artificial learning agent, we are able not only to study the collective properties that arise from the given tasks, but also to analyze the individual responses that agents learn and that, in turn, lead to the swarm formation. Thus, we observe for instance that the tendency to align with the neighbors in the $d_F=21$ case increases with the density of neighbors surrounding the agent. In the case of a training with $d_F=4$, we observe that the individuals tend to move to the region with higher number of neighbors, which leads to high cohesion at the collective level.

We note that the task faced by our artificial agents, of reaching a food source, is closely related to the behaviors studied in the context of foraging theory. For this reason, we compare the individual trajectories that result from the learning process to the principal theoretical models in that field. We show that most of the individual trajectories resulting from the training with distant resources ---which leads to strongly aligned swarms--- are best fitted by composite correlated random walks consisting of two modes, one intensive and one extensive, whose mean step lengths are $\hat{\lambda}_I^{-1}\simeq 2.7$ and $\hat{\lambda}_E^{-1}\simeq 75$, respectively. A smaller fraction of these trajectories is best fitted by power-law distributions with exponents $\hat{\mu}\cong 1.6$, that is, L\'evy walks. The exponent of the power-law distribution we obtain is close to 2, which is the optimal L\'evy walk for maximizing the rate of target encounters in environments with sparsely distributed, renewable resources \citep{Viswanathan99,Raposo09,Wosniack17}. Moreover, our results are in agreement with the study of Reynolds \citep{Reynolds13} that shows that animals can approximate L\'evy walks by adopting a composite correlated random walk.

In contrast, agents that were trained to find nearby resources and follow the dynamics of cohesive swarms present normal-diffusive, Brownian-like trajectories that do not exhibit two movement modes but just one. 

One crucial point of this analysis is that our simulated agents move in a multi-agent context and their movement patterns are therefore determined by the swarm dynamics they have developed through the learning process. In particular, we provide a new perspective and additional insight on the studies mentioned above regarding L\'evy walks and CCRW, since the individual trajectories that are best fit by these two models arise from a collective motion with very specific features such as strong alignment and decaying cohesion. This, together with the fact that the individual responses emerge as a result of the learning process, provides an example of how L\'evy-like trajectories can emerge from individual mechanisms that are not generated by a L\'evy walk process. In this sense, our work provides an unusual example to consider within the emergentist versus evolutionary debate on L\'evy walks (see e.g. \citep{Pyke15,Wosniack17}).

To conclude, we have applied a model of artificial agency (PS) to different foraging scenarios within the framework of collective motion. We have shown that, without any prior hard-wired interaction rules, the same agents develop different individual responses and collective interactions, depending on the distance they need to travel to reach a food source. Agents form strongly aligned swarms to stabilize their trajectories and reach distant resources, whereas they form cohesive, unaligned swarms when the resources are near. In addition, we have shown that L\'evy-like trajectories can be obtained from individual responses that do not have a simple theoretical model as the underlying process, but instead are generated and arise from the interplay of a fine-grained set of learned individual responses and the swarm behavior that emerges from them at a collective level.

This work provides a new framework for the study of collective behavior, which supports more detailed and realistic representations of individuals' sensory and motor abilities and different types of environmental pressures. It would be interesting to apply this approach to the more complex collective behaviors that arise in two- and three-dimensional environments. Furthermore, the PS model allows for a variety of new scenarios to explore in the context of behavioral biology, since different reward schemes can easily be implemented and studied.

\section*{Acknowledgements}
We would like to thank Oleksandr Chepizhko, Vladimir Palyulin, Gorka Muñoz Gil, Nicol\'as Tiz\'on Escamilla and Fernando Peruani for helpful discussions. This work was supported in part by the Austrian Science Fund (FWF) through the SFB BeyondC F71. HJB was also supported by the Ministerium f{\"u}r Wissenschaft, Forschung und Kunst Baden W{\"u}rttemberg (AZ:33-7533-30-10/41/1).

\newpage
\bibliography{SwarmFormation}

\appendix
\numberwithin{equation}{section} 

\section{Additional analysis}
\setcounter{equation}{0}
\renewcommand{\theequation}{A.\arabic{equation}}

In this section, we provide additional information about the dynamics presented in Section~\ref{learning diff scen} of the main text.

\subsection{Transition from cohesive to aligned dynamics}\label{APP transition}
\begin{figure}[htb!]
\includegraphics[width=3.4in]{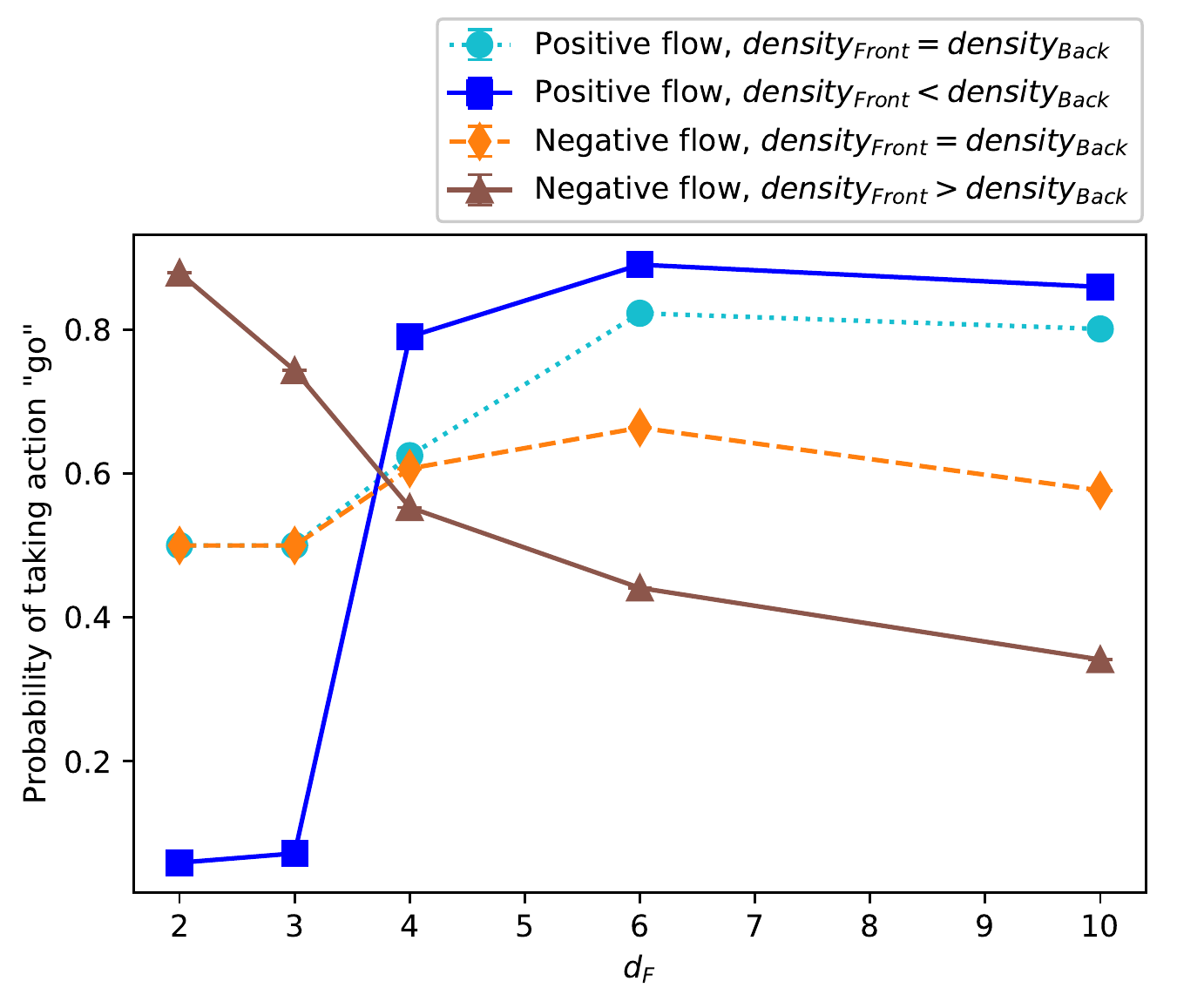}
\caption{Final probability of taking the action "go" depending on the learning task (increasing distance to food source $d_F$) for four significant percepts (see legend). Average is taken over one ensemble consisting of 60 agents.}\label{app fig phase tran}
\end{figure}

\begin{figure}[htb!]
\includegraphics[width=3.4in]{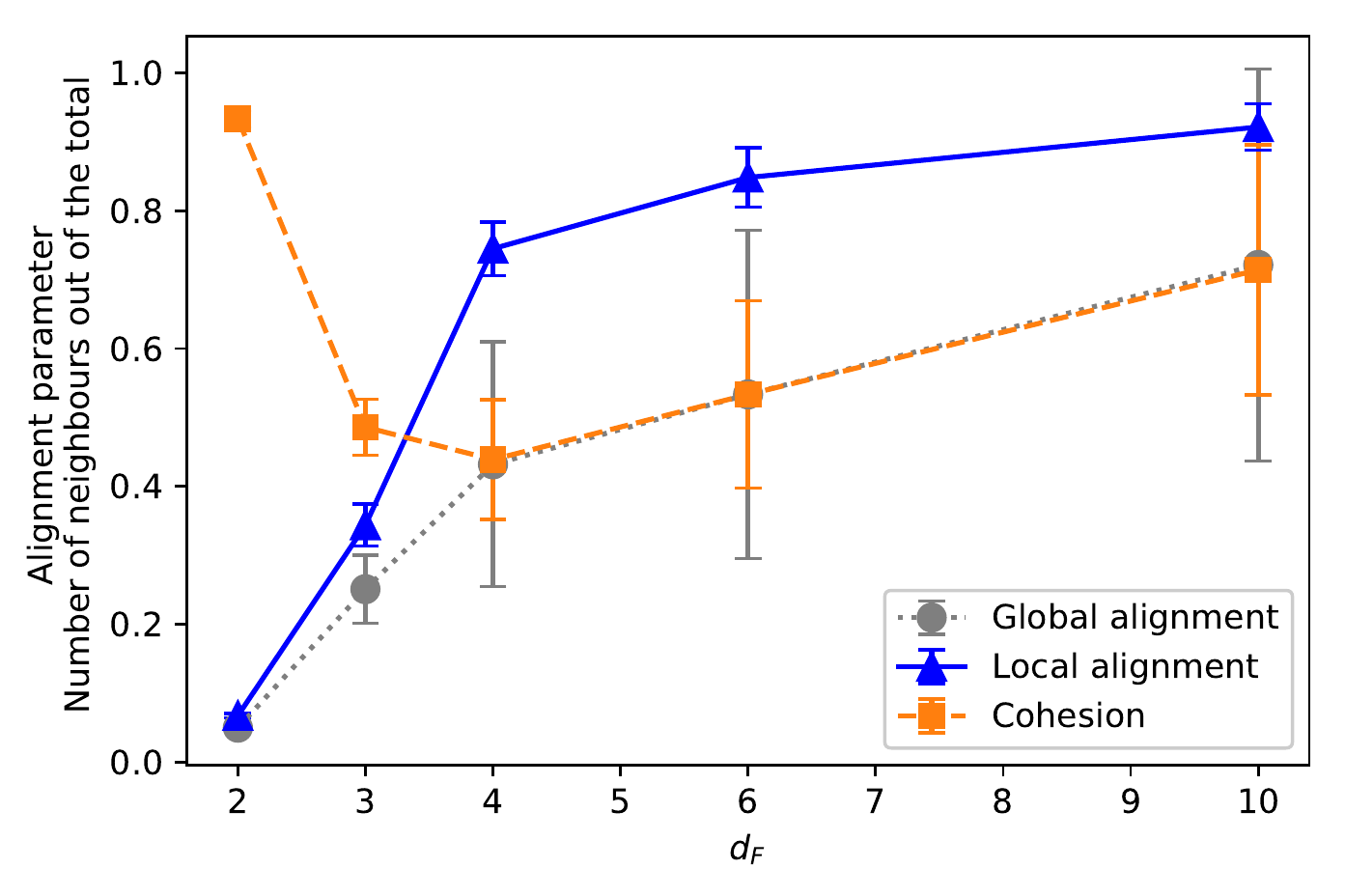}
\caption{Average number of neighbors (as a fraction of the total ensemble size), global and local alignment parameter as a function of the distance $d_F$ to the point where food is placed during training. Each point is the average of the corresponding parameter over all interaction rounds ($50$) of one trial, and then over $100$ trials. One trained ensemble of 60 agents is considered for each value of $d_F$.}\label{app fig phase tran 2}
\end{figure}

First, we analyze in detail why there is a transition at $d_F\simeq 6$ (see Figs.~\ref{comparison_matrixdf} and~\ref{comparison all df}) from the regime where cohesive swarms emerge to the regime where aligned swarms emerge as a result of the learning processes. We attribute this phenomenon to the fact that the agents are initialized in a region of size $2V_R$ (12 in our case), which means that a food source placed at $d_F=6$ is exactly at the edge of this region. Consider the case where the food is placed inside the initialization region: in this case, it is most likely that agents will find the food---which is the condition for being rewarded--- while they are surrounded by many neighbours. Consequently, behaviors that entail approaching or staying with other agents are more likely to lead to rewards ---effectively, agents learn to 'join the crowd'. However, if the food is placed outside the initial region, agents need to leave regions where the density of agents is high at the beginning of the trial, but they also need to stabilize their orientations, which is best achieved by aligning with one's neighbors. We have tested this hypothesis by changing the initial region. Figures~\ref{app fig phase tran} and \ref{app fig phase tran 2} show analogous data to Figures~\ref{comparison_matrixdf} and~\ref{comparison all df}, but with agents initialized in the first $V_R$ positions of the world (half of the previous region). We observe that the transition in behavior happens at $d_F=3$ in this case, which is the edge of the initial region.

\subsection{Details on analysis of alignment}\label{APP alignment}
\begin{figure}[htb!]
\subfigure[Agents trained with $d_F=21$]{\includegraphics[width=3.4 in]{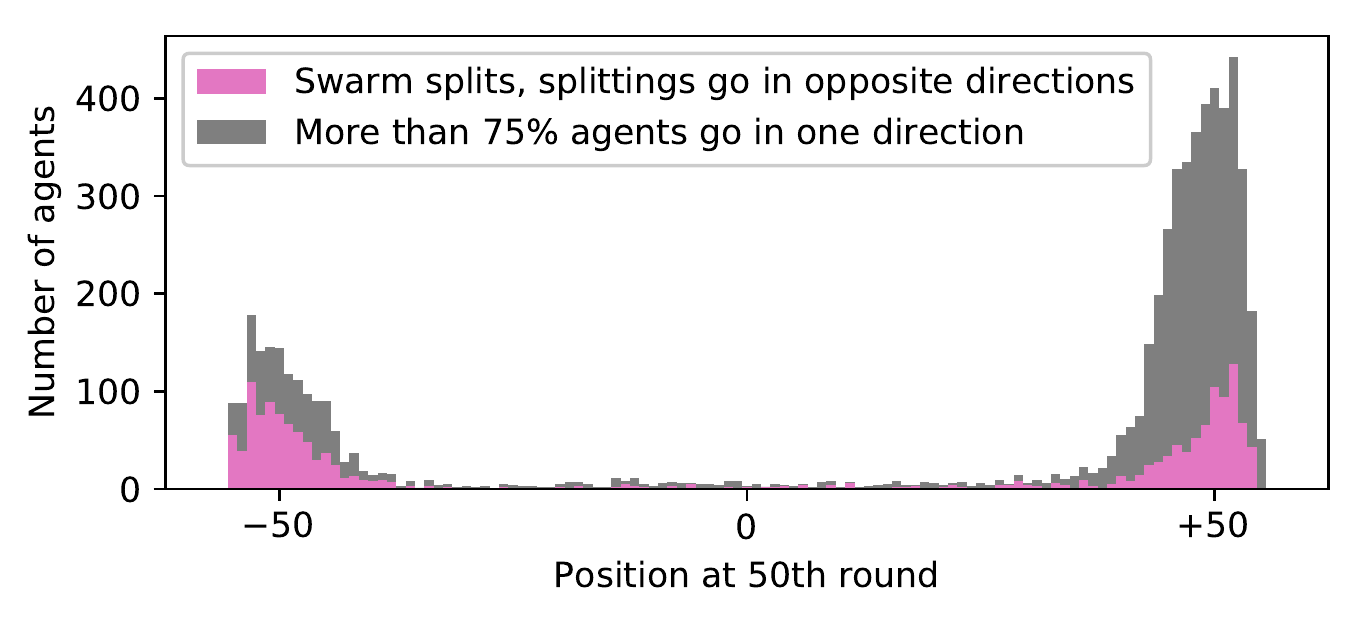}}
\subfigure[Agents trained with $d_F=6$]{\includegraphics[width=3.4 in]{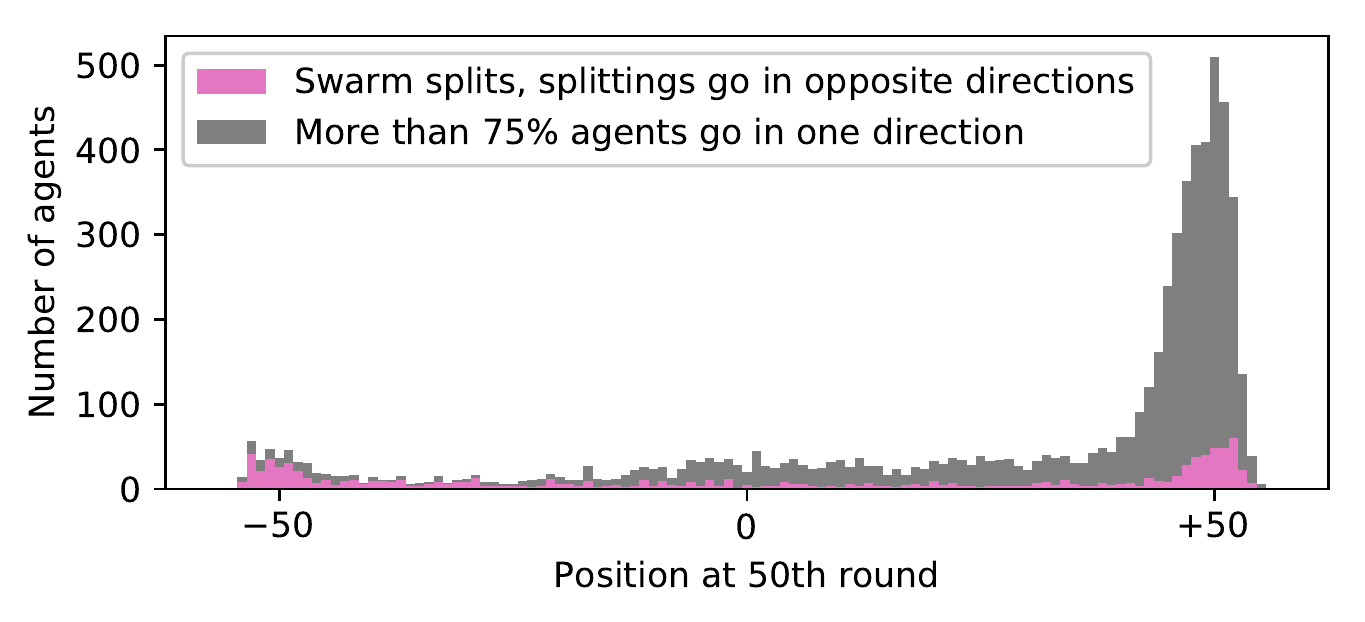}}
\subfigure[Agents trained with $d_F=4$]{\includegraphics[width=3.4 in]{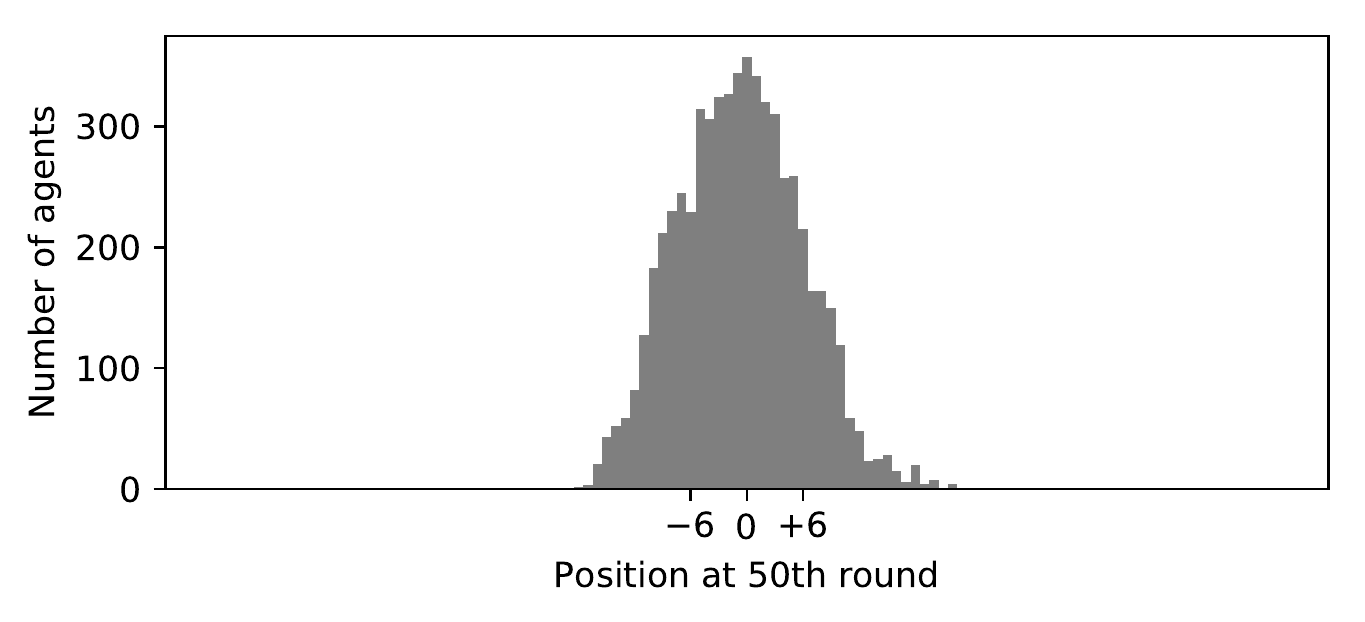}}
\caption{Stacked bar graph showing the number of agents that are located at a given position at the end of one trial (at the 50th interaction round). The graph is centered in $C$, which is the middle of the initial region (value 0 in the horizontal axis). Each data set (for each trial) is processed such that the majority of agents travel to the positive side of the horizontal axis. 100 trials of (already trained) ensembles of 60 agents are considered (one ensemble per trial). (a) Out of the 100 ensembles, 72 travel as one big swarm (grey) and 28 split into two subswarms that go in opposite directions (pink). In order to show that these are complementary subsets of the data, grey bars are \textit{stacked} on top of pink bars. (b) Out of the 100 ensembles, 83 travel as one big swarm (grey) and 17 split in two subswarms that go in opposite directions (pink). (c) All ensembles are strongly cohesive and do not split. Agents do not travel beyond the initial region (marked in the horizontal axis).}\label{splitting}
\end{figure}

In this section, we elaborate more on the splitting of the swarm that we observe in some of the trials for training with $d_F=21$. In order to study this, we perform a simulation of 100 trials with ensembles of agents that are already trained with $d_F=21$. Figure~\ref{splitting} (a) shows that, in some of these trials, almost all agents form one big swarm\footnote{We take the threshold for 'a single swarm' to be that $75\%$ of agents move in the same direction.} ($\phi\simeq 0.85$) that goes in one direction, with few agents moving away from the swarm (grey histogram), whereas in other trials they form two swarms ($\phi\simeq 0.55$), roughly of similar size, that travel in opposite directions (pink histogram). Locally, agents are strongly aligned, as can be seen in Fig.~\ref{comparison all df}, where average local alignment parameter reaches 0.9 for $d_F=21$. For $d_F=6$, the swarm behavior is similar to the one observed for $d_F=21$ (see Fig.~\ref{splitting} (b)), but the local alignment is not so strong, so there are more agents that go out of the swarm. For swarms trained with $d_F=4$ (Fig.~\ref{splitting} (c)), we observe that there is no splitting and agents do not move beyond the initial region.

\subsection{Details on analysis of cohesion}\label{APP cohesion}
\begin{figure}[htb!]
\includegraphics[width=3.4in]{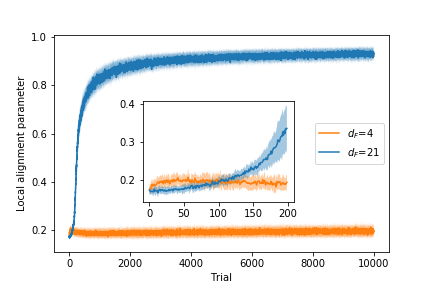}
\caption{Evolution of the local alignment parameter through the learning processes with $d_F=4,21$. Each point is the average of the corresponding parameter over all interaction rounds ($50$) of one trial. 20 independently trained ensembles of 60 agents each are considered for the average.}\label{app fig local evol}
\end{figure}

In this section, we provide an additional plot (Fig.~\ref{app fig local evol}) of the evolution of the local alignment parameter through the learning process for $d_F=4,21$. We observe that the increase of the local alignment parameter from trial 100 to trial 200 is the reason why the average number of neighbors decays at these same trials in Fig.~\ref{evol cohesion} (see inset). At these trials, agents have not yet learned to form swarms, but some of them have learned to go straight and started to learn to align with their neighbors. Thus, these agents are already able to go away from the initial region where the rest of agents are still doing a random walk. Consequently, these agents in particular have fewer neighbors, which reduces the overall average number of neighbors. For higher values of the local alignment parameter, as seen from trial 200 onwards, agents start to form strongly aligned swarms, which increases cohesion and consequently the number of neighbours $M$. 

\section{Statistical methods}\label{appendix stat}
\setcounter{equation}{0}
\renewcommand{\theequation}{B.\arabic{equation}}

The statistical methods we have applied in Sec.~\ref{stat results} of the main text are explained and detailed in this section. The statistical analysis consists of three steps: (i) the MLE estimation of the model parameters, for all models; (ii) the best model selection by means of the AIC and (iii) the absolute fit tests for the best model.

\subsection{Likelihood functions and MLE}\label{appendix mle}
The expression of the likelihood function~\eqref{likelihood function} can lead to computational underflows when the sample size is large\footnote{Large product of terms below one gives values close to zero.}, so the usual procedure is to maximize its logarithm instead (which leads to the same result since the function~\eqref{likelihood function} is monotonic). For i.i.d. variables, the log-likelihood function has the simple expression,
\begin{equation}
\ln \mathcal{L}(\theta|\ell_{i=1..S})=\sum_{i=1}^{S} \ln p\,(\ell_i,\theta),\label{loglikelihood fun}
\end{equation}
where $S$ is the sample size and $p(\ell_i,\theta)$ is the PDF of the given model ---that depends on the model parameters $\theta$--- evaluated at the data point $\ell_i$.

The first three models have i.i.d. variables, so the computation of their log-likelihood functions is straightforward once the PDFs of each model are defined (eqs.~\eqref{eq BW}, \eqref{eq CRW}, \eqref{eq PL}). However, the log-likelihood function of the CCRW model cannot be expressed as a sum of the logarithms of the PDFs evaluated at each data point. From eq.~\eqref{likelihood function}, the expression in the case of the CCRW can be written as,
\begin{multline}\label{likelihood fun CCRW}
\mathcal{L}(\delta,\lambda_I,\lambda_E,\gamma_{II},\gamma_{EE}|\ell_{i=1..S})= \\
=\delta_M P(\ell_1) \prod_{i=2}^S \Gamma P(\ell_i) \textbf{1},
\end{multline}
where,
\begin{align}
&\delta_M=\begin{pmatrix} \delta & 1-\delta \end{pmatrix},  \label{eq delta M}\\
&P(\ell)=\begin{pmatrix}
p_I(\ell) & 0 \\
0 & p_E(\ell)
\end{pmatrix}, \label{eq Px}\\
&\Gamma =\begin{pmatrix}
\gamma_{II} & 1-\gamma_{II} \\
1-\gamma_{EE} & \gamma_{EE}
\end{pmatrix}, \label{eq Gamma} \\
&\textbf{1}=\begin{pmatrix} 1 \\ 1 \end{pmatrix}, \label{eq id}
\end{align}
and $p_I(\ell)$ and $p_E(\ell)$ are given in eqs.~\eqref{eq CCRW pi} and~\eqref{eq CCRW pe} (note that they depend on $\lambda_I$ and $\lambda_E$, respectively). Since the variables are not independent in this case, the log-likelihood function cannot be directly obtained with expression~\eqref{loglikelihood fun}. In addition, function~\eqref{likelihood fun CCRW} cannot be directly computed due to underflow errors. To avoid this, we apply the techniques explained in chapter 3 of \citep{Zucchini09} (specifically, the algorithm for the computation of the log-likelihood function given in appendix A.1.3 of \citep{Zucchini09}).

Once the log-likelihood functions are computed, the maximization (minimization of the negative log-likelihood function) with respect to the model parameters is performed using the Python function scipy.optimize.minimize. The MLE parameters obtained for each model are given in Table~\ref{tabla:mle par}. The MLE of the minimum step length can be directly considered to be the observed one \citep{Edwards12} (in our case it is $\ell=1$ since agents move one position per interaction round).

\subsection{Goodness-of-fit tests}\label{gof section}
In this work, we have performed two types of goodness-of-fit (GOF) tests; one for the models with i.i.d. variables (BW, CRW and PL) and a different one to account for the temporal autocorrelation of the CCRW model. For the BW, CRW and PL models, we apply a likelihood ratio test to compare the likelihood of the observed frequencies to the likelihood of the theoretical distribution that corresponds to the given model. More specifically, we compute the log-ratio\citep{Clauset09},
\begin{equation}
\mathcal{R}=\sum_{i=1}^S [\ln f_{obs} (\ell_i)-\ln f_{th} (\ell_i)],
\end{equation}
where $S$ is the sample size and $f_{obs}$, $f_{th}$ are the observed and theoretical frequencies of the $i$th step length, respectively. Note that the theoretical frequency is just the probability (eq.~\eqref{eq BW}, eq.~\eqref{eq CRW} or eq.~\eqref{eq PL} depending on the analyzed model) of the $i$th step length times the sample size $S$. 

Normally, likelihood ratios like $\mathcal{R}$ above are used to compare two competing theoretical models, in which case a large absolute value of $\mathcal{R}$ indicates that one model is clearly better than the other. In order to assess how much better it is, one asks how likely it is that a given absolute value of $\mathcal{R}$ could have arisen purely from chance fluctuations, if in fact both models were equally good. This is quantified by the \emph{$p$-value} (App. C, eq. (C.6) of ref.~\citep{Clauset09}). When one compares two theoretical models, finds a large $|\mathcal{R}|$ and its corresponding $p$-value is small, this indicates that the value $\mathcal{R}$ is unlikely to be a chance fluctuation, and that one can therefore exclude one model with high confidence. 

\begin{figure}[htb!]
\includegraphics[width=3.4in]{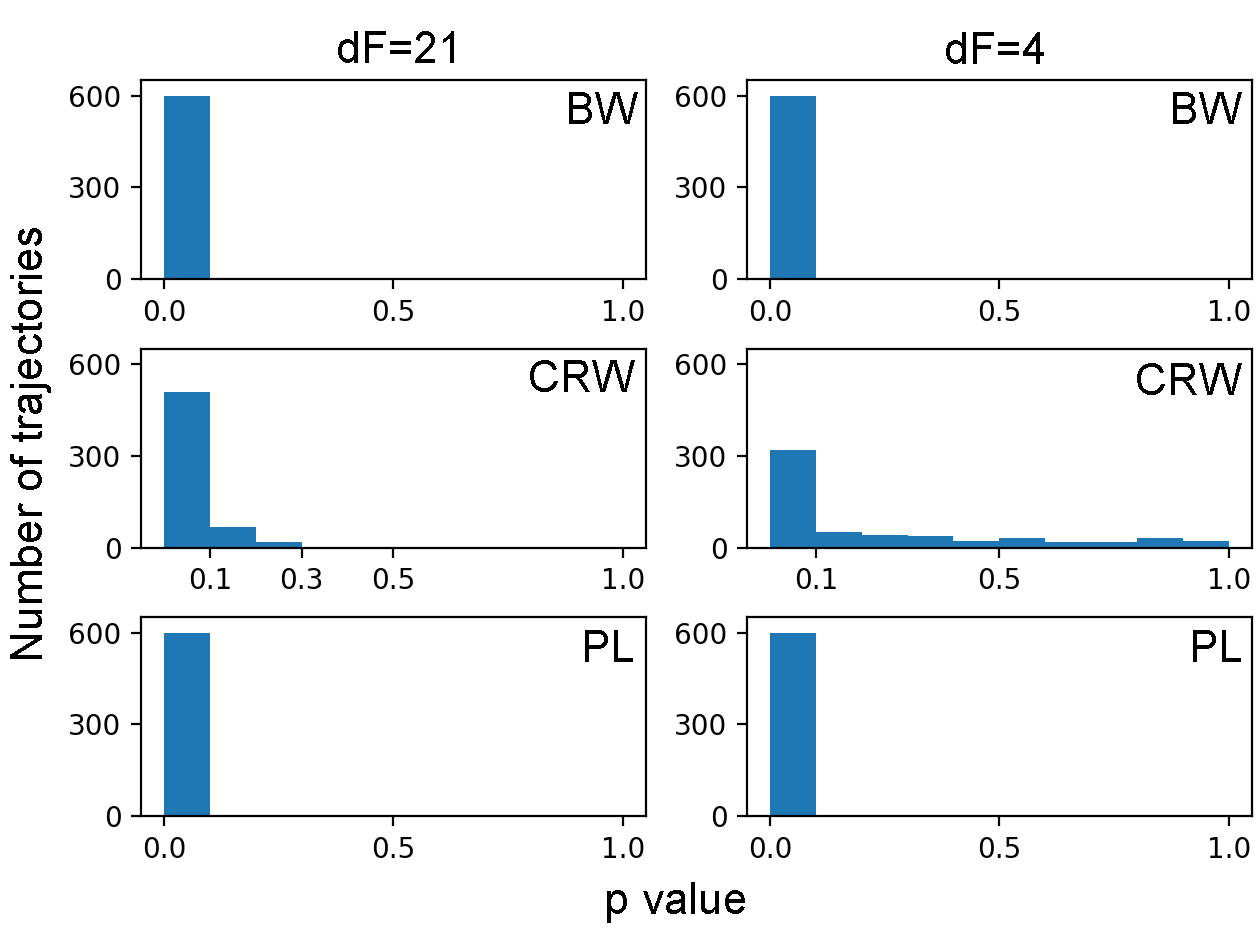}
\caption{Histograms of the $p$-values obtained for the BW, CRW and PL models in the $d_F=21$ and $d_F=4$ cases. In our goodness-of-fit test, $p$-values close to zero rule out the proposed theoretical model, while values close to 1 represent compatibility with the model.}\label{pvalues fig}
\end{figure}

In our case, however, a good fit between the theoretical model and the observed frequencies manifests as small $|\mathcal{R}|$ and correspondingly large $p$. Small $p$-values, on the other hand, indicate that it is unlikely that the data were generated by the proposed model. One can therefore interpret $1-p$ as the probability with which we can rule out the proposed theoretical model. The $p$-values obtained in our analysis are given in Fig.~\ref{pvalues fig}.

In the case of the CCRW model, one cannot directly perform a GOF test on the raw data points due to the autocorrelation present in the HMM model.

We circumvent this problem using \emph{pseudo-residuals}, as described in \citep{Zucchini09}. Given a continuous random variable $X$ and a function $F(X)$ defined by the cumulative distribution function (CDF),
\begin{equation}
F(X=x) := Pr(X\leq x),
\end{equation}
the pseudo-residual $u$ is obtained by sampling a value $x$ of $X$, then taking the corresponding value of the function $F$. If $X$ is sampled from some probability distribution $P_{exp}$ and we take $F_{exp}$ to be the CDF of that same distribution,
\begin{equation}
F_{exp}(X=x) = \int^x P_{exp}(X=x')dx',
\end{equation}
then one can show that the resulting probability distribution over the pseudo-residuals is in fact uniform $U(0,1)$ \citep{Zucchini09}. If, on the other hand, we take $F_{theo}$ to be the CDF derived from some proposed theoretical distribution $P_{theo}$, then the pseudo-residuals will in general not be uniformly distributed. By testing whether the pseudo-residuals with respect to a given theoretical model are uniformly distributed, one can therefore test whether the model is a good fit for the data. 

In order to accommodate discrete variables, one introduces so-called mid-pseudo-residuals,
\begin{equation}
u^m = (u+u^-)/2,
\end{equation}
where $u$ is obtained by sampling a value $x$ of $X$ and taking the corresponding $F(X=x)$, as above, while $u^- = F(X=x^-)$ is the value of $F$ at the greatest possible realization that is strictly less than the sampled $x$.

Our data consists of a time-series of step lengths $\ell_t$, each of which gives rise to one mid-pseudo-residual $u^m_t$. Therefore, the first step length is denoted $\ell_1$ and the last one $\ell_S$, since $S$ is the sample size. In order to be consistent with notation in Sec.~\ref{SEC Analysis of the learned dynamics} for step lengths, we use in the following the upper case $L$ to denote the random variable and the lower case $\ell$ to denote one realization of it.

Crucially, the probability distribution over step lengths at each time-step is different, since it is correlated with the lengths of preceding steps:
\begin{align}
u_t^-&=\text{Pr}(L_t<\ell_t|L^{(-t)}=\ell^{(-t)}),\\
u_t&=\text{Pr}(L_t\leq \ell_t|L^{(-t)}=\ell^{(-t)}),
\end{align}
where the expression for the conditional probability (\citep{Zucchini09}, (Chapter 5)) is in our case,
\begin{multline}
\text{Pr}(L_t\leq \ell|L^{(-t)}=\ell^{(-t)})= \\
=\frac{\delta_M P(\ell_1) B_2...B_{t-1} \Gamma Q(\ell) B_{t+1}...B_T \textbf{1}}{\delta_M P(\ell_1) B_2...B_{t-1} \Gamma B_{t+1}...B_T \textbf{1}},
\end{multline}
where $\delta_M$, $P(\ell)$, $\Gamma$ and $\textbf{1}$ are defined in eqs. \eqref{eq delta M}, \eqref{eq Px}, \eqref{eq Gamma} and \eqref{eq id} respectively and,
\begin{align}
&B_t=\Gamma P(\ell_t),\\
&Q(\ell)=\begin{pmatrix}
q_I(\ell) & 0 \\
0 & q_E(\ell)
\end{pmatrix},
\end{align}
where $p_I(\ell)$ and $p_E(\ell)$ are the PDFs defined in eq.~\eqref{eq CCRW pi} and~\eqref{eq CCRW pe} and $q_I(\ell)$ and $q_E(\ell)$ are their corresponding CDFs, respectively. Note that, in this expression, the parameters of the model are fixed (MLE parameters). Again in this case, a rescaling is needed in order to avoid underflows in the computation (see algorithm in App. A.2.9 of ref.~\citep{Zucchini09}). 

\begin{figure}[htb!]
\includegraphics[width=3.4in]{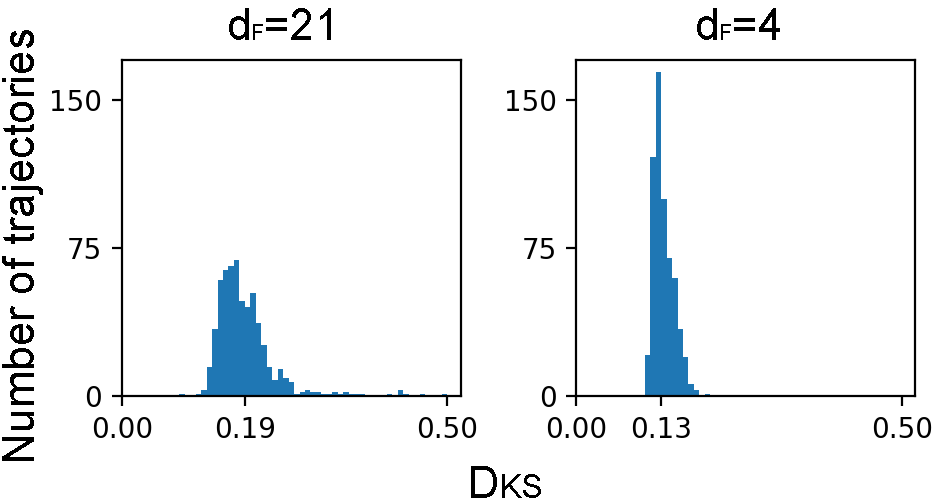}
\caption{Histograms of the $D_{KS}$ distances obtained in the GOF test of the CCRW model, for (left) $d_F=21$ and (right) $d_F=4$.}\label{DKS}
\end{figure}

In summary, we first compute the mid-pseudo-residual for each data point and then we perform a GOF test on them. Since the probability distribution of the mid-pseudo-residuals approaches that of a continuous variable, one can apply a Kolmogorov-Smirnov (KS) test to check for uniformity. The KS statistic computes the distance ($D_{KS}$) between the CDF of the empirical data (in this case, the values $u_t^m$) and the CDF of the reference distribution (in this case, $U(0,1)$). Therefore, a value $D_{KS}=0$ means that the data is distributed exactly as the reference distribution. The maximum KS distance is $D_{KS}=1$. One obtains one value of $D_{KS}$ for each individual trajectory (we perform the analysis on 600 trajectories for each type of swarm dynamics). The average value of the KS distance that we have obtained is $D_{KS}=0.189\pm 0.046$ for the trajectories of agents trained with $d_F=21$ and $D_{KS}=0.134\pm 0.016$ for the ones of agents trained with $d_F=4$. All the values of $D_{KS}$ are displayed in a histogram form in Fig.~\ref{DKS}.

\subsection{Tables}\label{APP tables}
\begin{figure}[b!]
\includegraphics[width=3.4in]{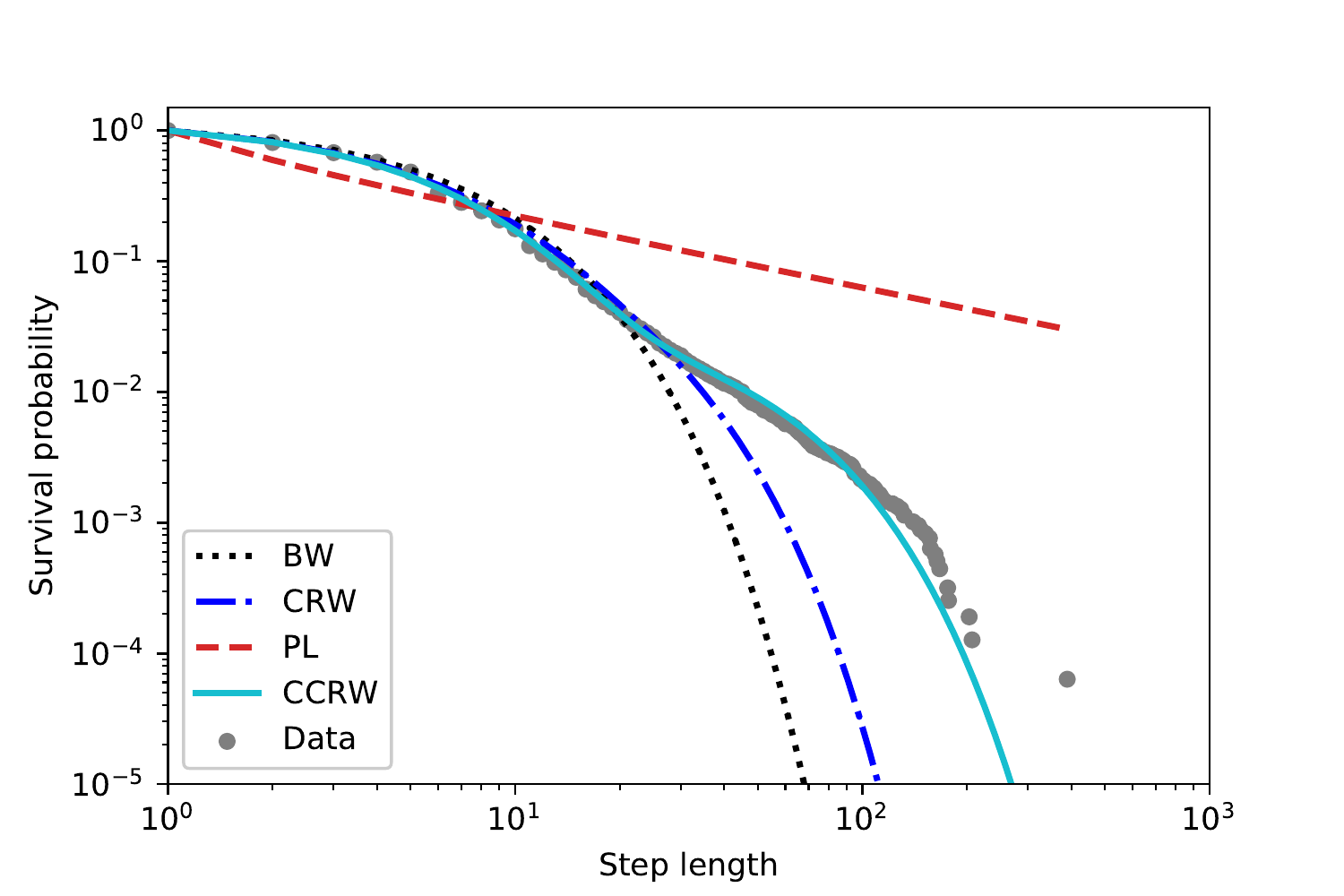}
\caption{Survival probability (cumulative percentage of step lengths larger than the corresponding value in the horizontal axis) as a function of the step length. Trajectory of one agent trained with $d_F=21$, which has an Akaike value of 1 for the CCRW model. This individual was chosen for achieving the closest fit to the CCRW model of all agents trained with $d_F=21$. The survival distributions of the four candidate models are also plotted. The distributions for each model are obtained considering the MLE parameters. }\label{app survival best ccrw}
\end{figure}

\begin{figure}[htb!]
\includegraphics[width=3.4in]{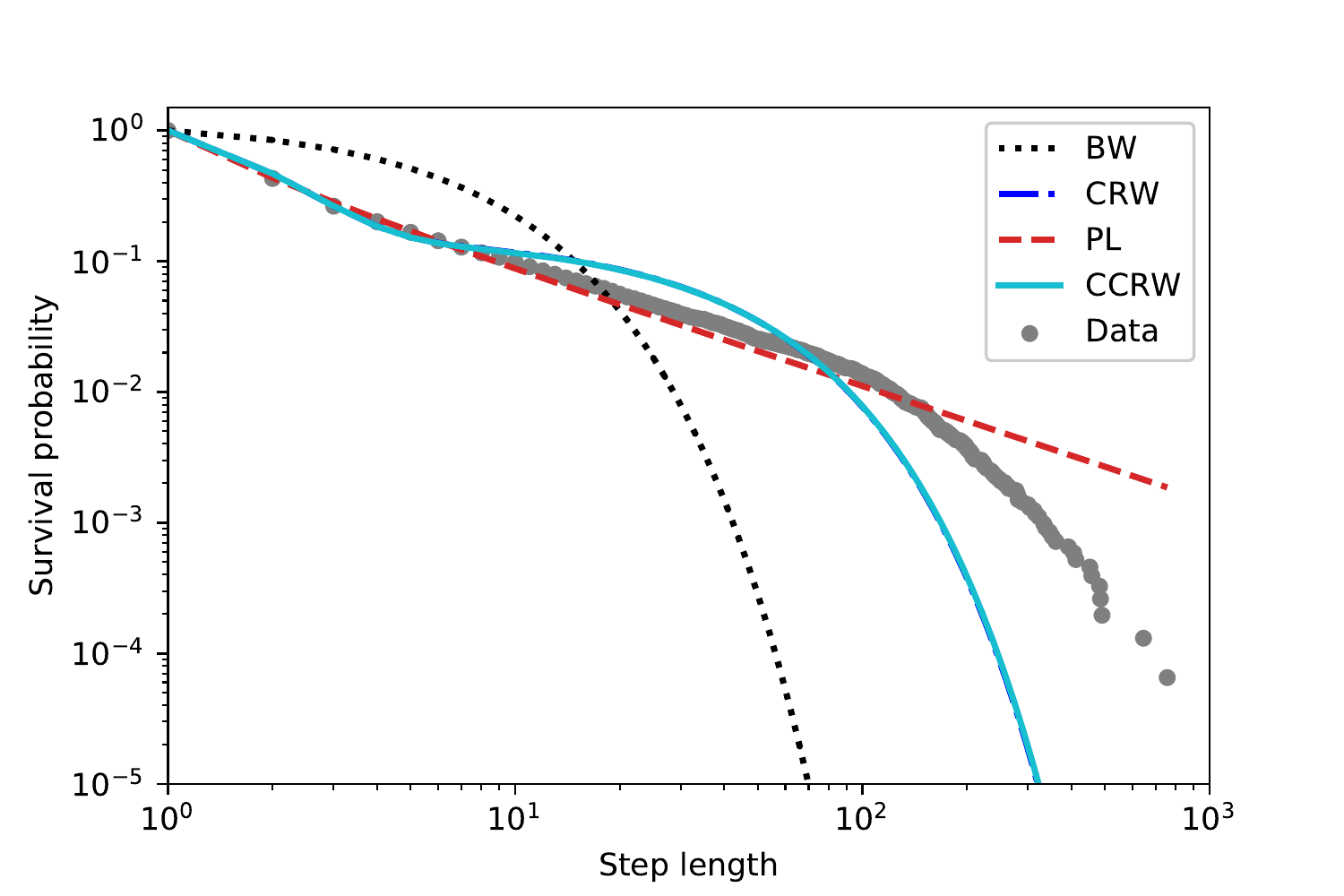}
\caption{Survival probability (cumulative percentage of step lengths larger than the corresponding value in the horizontal axis) as a function of the step length. Trajectory of one agent trained with $d_F=21$, which has an Akaike value of 1 for the PL model. This individual was chosen for achieving the closest fit to the PL model of all agents trained with $d_F=21$. The survival distributions of the four candidate models are also plotted. The distributions for each model are obtained considering the MLE parameters. }\label{app survival plot}
\end{figure}

Examples of the results of the statistical analysis for one trajectory are given in Tables~\ref{tabla:likelihood and par 4}, \ref{tabla:likelihood and par}, and \ref{tabla:likelihood and par LW}. The trajectories considered correspond to the ones displayed in figures~\ref{survival} (b), \ref{app survival best ccrw} and \ref{app survival plot}, respectively. In addition, figures \ref{app survival best ccrw} and \ref{app survival plot} provide the survival distributions of the trajectories that have the best goodness-of-fit parameter for the CCRW and the PL models, respectively.

\begin{table}[htb]
\begin{tabular}{|c|c|c|c|c|c|}
\hline
Model & k & $AIC$ & $\Delta_i$ & $w_i$ & $p$-value\\ \hline \hline
BW & 2 & 130534.03 & 0 & 0.87 & 0.0018 \\ \hline 
CRW & 4 & 130538.03 & 4 & 0.12 & 0.96\\ \hline
PL & 2 & 144670.67 & 14136.64 & 0 & $<0.01$\\ \hline 
CCRW & 6 & 130542.03 & 8 & 0.01 & $D_{KS}=0.18$\\ \hline 
\end{tabular}
\caption{Results of the statistical analysis of the trajectory from Fig.~\ref{survival} (b). This individual was chosen for achieving the closest fit to the BW and CRW models of all agents trained with $d_F=4$.}
\label{tabla:likelihood and par 4}
\end{table}

\begin{table}[htb!]
\begin{tabular}{|c|c|c|c|c|c|}
\hline
Model & k & $AIC$ & $\Delta_i$ & $w_i$ & $p$-value \\ \hline \hline
BW & 2 & 87207.06 & 2104.71 & 0 & $<0.01$\\ \hline 
CRW & 4 & 85676.13 & 573.78 & 0 & $<0.01$\\ \hline
PL & 2 & 94815.86 & 9713.51 & 0 & $<0.01$\\ \hline 
CCRW & 6 & 85102.35 & 0 & 1 & $D_{KS}=0.094$\\ \hline 
\end{tabular}
\caption{Results of the statistical analysis of the trajectory from Fig.~\ref{app survival best ccrw}. This individual was chosen for achieving the closest fit to the CCRW model of all agents trained with $d_F=21$.}
\label{tabla:likelihood and par}
\vspace{0.3in}
\begin{tabular}{|c|c|c|c|c|c|}
\hline
Model & k & $AIC$ & $\Delta_i$ & $w_i$ & $p$-value \\ \hline \hline
BW & 2 & 85615.24 & 28236.12 & 0 & $<0.01$\\ \hline 
CRW & 4 & 58473.79 & 1094.67 & 0 & $<0.01$\\ \hline
PL & 2 & 57379.12 & 0 & 1 & $<0.01$\\ \hline 
CCRW & 6 & 58471.36 & 1092.24 & 0 & $D_{KS}=0.29$\\ \hline 
\end{tabular}
\caption{Results of the statistical analysis of the trajectory from Fig.~\ref{app survival plot}. This individual was chosen for achieving the closest fit to the PL model of all agents trained with $d_F=21$.}
\label{tabla:likelihood and par LW}
\end{table}

\newpage

\end{document}